\documentclass[11pt,a4paper]{article}
\pdfoutput=1

\usepackage{cite}
\usepackage[english]{babel}
\usepackage[utf8]{inputenc}
\usepackage{amsmath}
\usepackage{graphicx}
\usepackage[nottoc]{tocbibind}
\usepackage{amssymb}

\usepackage[colorlinks=true, allcolors=blue]{hyperref}
\usepackage{float}
\usepackage{multirow}
\usepackage{makeidx}

\numberwithin{figure}{section}
\numberwithin{equation}{section}

\newcommand{\be}{\begin{equation}}
\newcommand{\ee}{\end{equation}}
\newcommand{\bea}{\begin{eqnarray}}
\newcommand{\eea}{\end{eqnarray}}
\def\beal#1\eeal{\begin{align}#1\end{align}}   
\def\besp#1\eesp{\begin{multline}#1\end{multline}} 

\newcommand{\eqrf}[1]{}

\newcommand\ie{\textit{i.e.}\ }
\newcommand\eg{\textit{e.g.}\ }
\newcommand\cf{\textit{cf.}\ }

\newcommand{\aka}{{a.k.a.}\ }
\newcommand{\etc}{{\it etc.}\ }
\newcommand{\viz}{{\it viz.}\ }

\newcommand{\nn}{\nonumber}

\newcommand{\ph}{\varphi}

\newcommand{\La}{\Lambda}

\usepackage[dvipsnames]{xcolor}

\title{Asymptotic behaviour of the derivative expansion in the ERG}

\author{Tim R. Morris\\
\small \texttt{T.R.Morris@soton.ac.uk}\\
 \small \textit{STAG Research Centre, Department of Physics and Astronomy,}\\
  \small \textit{University of Southampton, Highfield, Southampton, SO17 1BJ, U.K.}
  }

\date{\today}
\makeindex
\begin{document}

\maketitle
\begin{abstract}
We show that the derivative expansion of the exact (functional) renormalization group is a divergent series in any dimension, both for an exponential cutoff and more general smooth cutoffs. We prove this by showing that within massless $\lambda\ph^4$ perturbation theory, such divergences arise first at two loops. From several lines of theoretical argument and by analysing infinite classes of two- and three-loop contributions, we conclude that the derivative expansion is an asymptotic series that initially converges towards the exact result before divergent behaviour takes over.
\end{abstract}
\newpage
\tableofcontents
\newpage

\section{Introduction}
\label{sec:intro}

The exact renormalization group (ERG) \cite{Wilson:1973}, also known as the functional renormalization group (FRG), has become one of the principal non-perturbative approaches to quantum and statistical field theory. Since the introduction of the functional flow equation \cite{Nicoll1977,Wetterich:1992,Morris:1993qb} for the Legendre effective (average) action $\Gamma[\ph]$, the method has found applications ranging from critical phenomena and phase transitions to gauge theories and quantum gravity \cite{Dupuis:2020fhh}. 

In practice, however, exact solutions of the flow equation are rarely possible and one must rely on approximation schemes. Among these, the derivative expansion has emerged as one of the most widely used and successful approximations \cite{Morris:1994ie,Morris:1994jc,Morris:1996xq,Morris:1997xj,Morris:1999ba,Litim:2001,Canet:2002gs,Canet:2003qd,Balog:2019rrg,DePolsi:2020pjk,Delamotte:2024xhn}. It has been demonstrated in a wide variety of systems, where it has proved capable of producing remarkably accurate estimates of, for example, critical exponents and scaling functions, in quantitative agreement with other high accuracy methods such as experiment, high-order perturbative methods, Monte Carlo simulations and conformal bootstrap calculations. For introductions and reviews see \cite{Morris:1994au,Morris:1996kn,Berges:2000ew,Morris:2000hm,Delamotte:2007pf,Dupuis:2020fhh}, and for recent applications see for example \cite{Bonanno:2026mzs,Maitiniyazi:2025pou,DePolsi:2025gqj,Cabrera:2024rgy,Skokov:2024fac,Chlebicki:2024qsh,CancinoArancibia:2024joz,Balog:2025,Baldazzi:2023pep,Delamotte:2024xhn}.
Derivative expansion approximations are defined by expanding in powers of derivatives of the fields up to some order, $O(\partial^{2n})$, discarding higher order contributions. In practice it is found that each successive order shows rapid convergence. The current state of the art is  $O(\partial^6)$ \cite{Balog:2019rrg}. 
However there is no complete understanding for why these approximations are so successful. 

It is known that the derivative expansion is exact in various limits, in particular the form of the large field behaviour is correctly reproduced by the expansion \cite{Morris:1994au,Morris:1996xq,Morris:1994jc,Morris:1998};  the lowest order $O(\partial^0)$, \aka the local potential approximation \cite{Nicoll:1974zz}, is in a sense exact for $O(N)$ scalar field theory in the large $N$ limit \cite{DAttanasio:1997he}; and the one-loop perturbative result is incorporated exactly \cite{Wetterich:1992}. Since the derivative expansion furnishes results that interpolate between these exact limits, some accuracy is then guaranteed. But the actual  accuracy achievable  points to some deeper mechanism. 

In this respect we note that, for a convenient choice of cutoff function, the two-loop $\beta$ function in massless four-dimensional $\lambda\ph^4$ theory can be computed to all orders in the derivative expansion, and it is found that the resulting expansion converges to the known exact answer \cite{Morris:1999ba}, see also \cite{Papenbrock:1994kf}. On the other hand, suggestive arguments for convergence have been proposed based on comparing the effective cutoff scale $\La$ to that of a mass $m\sim \La$ in the final (\ie cutoff-free) $\lambda\ph^4$ theory, whose self-energy 
\be \label{selfe}\Gamma(p,-p)=\frac{\delta^2\Gamma}{\delta\ph(p)\delta\ph(-p)}\ee 
would thus (by unitarity) have a radius of convergence $p^2/m^2=4$ or $9$, depending on whether it is in the broken or symmetric phase respectively \cite{Balog:2019rrg,DePolsi:2020pjk,Dupuis:2020fhh}.

In this paper we will make significant advances in understanding the mechanism behind convergence. We start with the following observation. If the derivative expansion truly converges, then it must also do so in the loop expansion, since this is nothing but a Taylor expansion of the full non-perturbative result in Planck's constant $\hbar$. The great benefit for us of working within the loop expansion, is that developing the derivative expansion to infinite order is then tractable. As mentioned above, this programme was initiated in ref. \cite{Morris:1999ba}, where convergence of the derivative expansion was confirmed in the examples considered. However, that analysis only went as far as the two-loop $\beta$ function in four-dimensional $\lambda\ph^4$ theory, leaving open the question of whether the convergence persists more generally. 

Here we report the result of pushing that analysis much further. Already at two loops, we find that the derivative expansion fails to converge for infinitely many vertices. It follows that the derivative expansion cannot converge non-perturbatively either. As a corollary, the very popular approximation technique of expanding the FRG in ever larger sets of local operators, also cannot converge, since this is just an expansion over a subset of all the operators treated in the derivative expansion.

Thus we are faced with the fact that in the published papers the derivative expansion gives increasingly accurate results at low $O(\partial^{2n})$ orders, and yet we know that eventually it must diverge as $n\to\infty$. Assuming that the successes up to now are not accidental, this  already tells us that the expansion must be \emph{asymptotic} in the sense that the series $\sum_{n=0} a_n$ has the property that successive partial sums approach the exact result up to some critical point $n=n_{cr}$, after which the terms $a_n$ stop decreasing and the series  diverges. 

Note that such a definition is not the Poincar\'e one. It is broader (and thus less powerful) than this. In the Poincar\'e definition  \cite{Dingle1973,OlverAsymptotics,bender1999advanced}, we need to have a tunable parameter, call it $\Delta$, that we can  tune, for example to zero, such that the terms in the series obey $a_{n+1}/a_n\to0$ as $\Delta\to0$, and such that in this limit the error in the partial sum vanishes faster than the last included term. Then provided $\Delta$ is small enough, the series will have the property we stated above that it converges initially up to some critical value $n=n_{cr}(\Delta)$, after which it diverges. 

In the derivative expansion there is no small parameter we can use to tune the $a_n$. In effect in the derivative expansion, this small parameter is set to $\Delta=1$. Indeed in sec. \ref{sec:watson},  we show, at the non-perturbative level, that the structure of the flow equation allows us to map its right-hand side into a form where Watson's lemma \cite{Watson:lemma,Miller:aaa} applies, guaranteeing that the derivative expansion is Poincar\'e asymptotic, except for the fact that the expansion parameter $g$ in this lemma is here fixed to $g=1$. Then it is a matter of luck whether the resulting series is asymptotic in the above sense. In the rest of the paper we gather multiple lines of theoretical evidence that strengthen the claim that we are indeed in the lucky regime such that the derivative expansion is an asymptotic expansion in that sense.

In particular in sec. \ref{sec:twoloops}, working in $d$ dimensions and with the same exponential cutoff, we identify the dominant two-loop contribution at large $O(\partial^{2n})$ and compute the leading part of this contribution in a large $n$ expansion. Its behaviour tells us a great deal about the derivative expansion at very high order. Indeed if, for a given operator, this leading part forms a convergent series, then at two loops the full derivative expansion for this operator is a convergent series. On the other hand, if the leading part for this operator forms a divergent series, then the full derivative expansion for this operator is a divergent series, both at two loops and non-perturbatively. 

What we actually find for this leading part is an oscillating series that converges for the two-point and four-point vertices in any dimension $d$, but for six-point vertices only in dimensions that are not too large, while for higher-point vertices the derivative expansion is divergent in any dimension. However their $O(\partial^{2n})$ expansion is such that the leading part converges to very high order, at least up to $O(\partial^{18})$ in important cases, before diverging exponentially fast in $n$. In sec. \ref{sec:asymptotics}, we prove that these leading terms form an asymptotic series in the above sense, in particular we prove that the most accurate result is achieved by truncating the series just short of the smallest term. For $2m$-point vertices, when $m$ is large, identifying which is the smallest term however requires going beyond the leading contributions, as we explain in sec. \ref{sec:eightpoint}. 

In sec. \ref{sec:threeloop} we identify a three-loop correction that we argue will give the dominant contribution overall, and compute its leading terms in the large $O(\partial^{2n})$ expansion. At this level, all vertices now have a divergent derivative expansion (and thus non-perturbatively the derivative expansion is divergent for all vertices) but we see again that the leading terms form an asymptotic expansion which converges this time up to $O(\partial^{10})$ (in important cases).

 Finally, in sec. \ref{sec:gen} we work with general smooth cutoffs and show again that the leading part of the  $O(\partial^{2n})$ expansion forms an asymptotic series in the above sense.

In more detail, the paper is structured as follows. After setting up the problem in secs. \ref{sec:setup} and \ref{sec:expc},  we review and begin extending the results of \cite{Morris:1999ba,Papenbrock:1994kf} in sec. \ref{sec:exp}. At one loop, derivative expansions of vertices are just Taylor expansions in $p_i/\La$, where the $p_i$ are their external momenta. With the choice of exponential cutoff, defined in sec. \ref{sec:expc}, these expansions have an infinite radius of convergence with the coefficients at $O(\partial^{2n})$ falling factorially with $n$, \cf sec. \ref{sec:oneloop}. 

That is no longer true at higher loops. The derivative expansion is then a numerical expansion, with no small parameter, just as it is non-perturbatively. In sec. \ref{sec:twoloop} we introduce $\Delta$ as a derivative expansion counting parameter so that the $O(\partial^{2n})$ contribution can be read off from the coefficient of $\Delta^n$. In reality $\Delta=1$, but its introduction helps in deriving identities, resumming series, and in discussing the rate of convergence. In particular we see in sec. \ref{sec:twoloop} that the two-loop self-energy, $\Gamma_2(p,-p)$, has a derivative expansion with radius of convergence $\Delta=2$. This means that the derivative expansion yields  an absolutely convergent numerical series, although the coefficients of its $p^{2r}$ part decay as $\sim n^r/2^n$, much slower than the factorial decay at one loop. On the other hand, once resummed, the coefficient of the $p^{2r}/\La^{2r}$ term has factorial decay in $r$. Thus, once resummed, the two-loop self-energy has a Taylor expansion in $p^2$ with an infinite radius of convergence, similar to vertices at one loop.

We then compute the two-loop four-point vertex at zero momentum via derivative expansion, using it in sec. \ref{sec:beta} to verify that the collected and resummed contributions yield the correct $\beta$ function.  This is the result of several contributions that all have convergent derivative expansions. In this section we find a simple analytical procedure for extracting the leading large $O(\partial^{2n})$ behaviour in such quantum corrections. In this way we show that whilst all the other contributions again have radius of convergence $\Delta=2$, the most complicated contribution converges more slowly, with $\Delta=3/2$. 

We explain the reason for this slowest convergence in sec. \ref{sec:anypoint}. It is the contribution that involves the highest-point vertex at one loop, namely the six-point vertex, and we show that this follows because the larger the mass dimension of the operator, the more slowly its derivative expansion converges at one loop. Then, in sec. \ref{sec:oneloops}, we go on to verify this by computing the leading $O(\partial^{2n})$ behaviour at one loop for all the $2m$-point vertices in any spacetime dimension $d$. 

In turn at higher loops, the poorest convergence arises in the contributions built on the most irrelevant operator at one loop. Continuing to work now in general spacetime dimension, and armed with that insight and our analytical procedure,  in sec. \ref{sec:twoloops} we compute the dominant contribution at large $n$ for any zero-momentum $2m$-point vertex at two loops, while also generalising this to derivative expansion contributions to $2m$-point $2r$-derivative operators (for all $r$). 

Turning to the convergence properties of these derivative expansions, we confirm the earlier result that the derivative expansions at two loops converge for two-point and four-point operators, but now we see that this is also true in any dimension $d$ and for operators containing any number $2r$ of derivatives. 

However already at the six-point level this is no longer the case: convergence for a $\partial^{2r}$ operator requires $d<10-2r$. Thus, for example, the zero-momentum $\ph^6$ vertex has a convergent derivative expansion only in $d<10$ dimensions, whilst the $\partial^8\ph^6$ operator (the derivatives being distributed amongst the fields in any order) does not have a convergent derivative expansion in any dimension $d\ge2$. 

Then for eight-point and higher-point operators, for any number $2r$ of derivatives, and in any dimension $d$, the derivative expansion \emph{does not converge}. However what we see in the leading behaviour is an oscillating series that behaves asymptotically in the sense described  above. The leading $O(\partial^{2n})$ contributions in fact continue to get smaller, thus yielding a convergent series, up to very high order, after which the series diverges exponentially fast.  In particular in sec.  \ref{sec:eightpoint} we see that amongst all the monomials $\ph^{2m}$, the leading part of the derivative expansion for the $\ph^{20}$ operator breaks down first (independent of dimension $d$) but for $d=3$ or $4$ dimensions, it nevertheless converges up to $O(\partial^{18})$, whilst for $d=2$ dimensions it converges up to $O(\partial^{20})$. See however the caveat below. All the other powers have series that continue to converge to even higher order, \eg $\ph^8$ has a series that converges up to $O(\partial^{48})$ in $d=3$ dimensions.

In sec. \ref{sec:asymptotics} we show that for these leading terms, considered as a formal series on their own, the natural definition of the exact result always lies between successive partial sums, thus proving that the leading terms form an asymptotic series in the sense described  earlier. In particular it follows that, barring numerical flukes, the smallest error will be achieved by truncating the series just short of the smallest term. 

We strengthen this evidence in sec. \ref{sec:watson}, by mapping the flow equation into a form where Watson's lemma  \cite{Watson:lemma,Miller:aaa} applies, as we already highlighted earlier. However, whilst the leading behaviour at large $O(\partial^{2n})$ is sufficient to decide whether the series converges or diverges in the limit $n\to\infty$, it is not necessarily sufficient to determine whether the full series behaves as an asymptotic series in the above sense, because that requires that subleading terms can still be neglected around $n\approx n_{cr}$. In sec. \ref{sec:eightpoint}, we check this for the dominant diagrammatic contribution to $\ph^8$ by comparing the leading terms to the exact result at large $O(\partial^{2n})$. We see that while the leading terms are roughly sufficient around $n_{cr}$ in that case (convergence actually breaking down at $O(\partial^{40})$), at higher $m$ and around $n\approx n_{cr}(m)$ we can no longer rely solely on the leading terms. In particular the results for $\ph^{20}$ (and for higher point vertices) quoted above are only true as a formal statement about the leading behaviour: subleading terms can be expected to contribute substantially.

So far we have been working with diagrammatic contributions to the $2m$-point vertices that are  built on $(2m\!+\!2)$-point one-loop vertices. We argued that these gave the leading contribution to the two-loop $2m$-point vertices at large $O(\partial^{2n})$ because they are built on the highest possible mass dimension one-loop vertex. In sec. \ref{sec:othertwo} we verify this by computing the leading $O(\partial^{2n})$ contribution from  $m-1$ further two-loop corrections to each vertex, built on one-loop four-point, six-point, and so on, up to $2m$-point vertices. We show that all these corrections do indeed give sub-leading contributions. In fact all those built on one-loop vertices with six or fewer legs have convergent derivative expansions.

In sec. \ref{sec:threeloop} we argue that the worst possible asymptotic behaviour at large $O(\partial^{2n})$ comes from a certain three-loop correction built on the leading two-loop contribution we computed in sec. \ref{sec:twoloops}. Computing the leading behaviour of this three-loop contribution, we find divergent expansions now for all vertices, however once again they are of asymptotic form. For example they imply that in $d=3$ dimensions, the derivative expansion for $\ph^2$ converges up to $O(\partial^{24})$, and for all monomials at least up to $O(\partial^{10})$. Once again however, the estimates for $n_{cr}(m)$ become increasingly unreliable for higher-point vertices, thus calculations that go beyond leading order are needed.

In sec. \ref{sec:otherthree} we strengthen the claim that the above three-loop correction provides the leading asymptotic behaviour overall, by demonstrating that derivative expansions that are split across different subdiagrams actually have better convergence. 

Then in sec. \ref{sec:gen} we work out what changes for other choices of cutoff than the exponential cutoff. These need to be smooth (differentiable to all orders) so that a derivative expansion exists to all orders. Choices in the literature, \eg those of ref. \cite{Balog:2019rrg}, lead to infrared regulated propagators $\Delta_{IR}(q,\La)$ with singularities in the complex $q^2$ plane a finite distance from the origin. The derivative expansion at one loop thus now has a finite radius of convergence. We show that this implies that at two loops the derivative expansion coefficients now all diverge as $n\to\infty$, growing factorially fast for all vertices. Nevertheless the leading terms still form an asymptotic expansion, in fact in the form usually found in the literature, and in a regime where accurate results can be extracted, provided the leading terms on their own are sufficient to determine the behaviour close to the minimum term. Furthermore for the leading contributions at large $O(\partial^{2n})$, we can again map the right-hand side of the flow equation into an integral that satisfies Watson's lemma (at $g=1$).

Finally, in sec. \ref{sec:conclusions} we discuss the significance of these results and draw our conclusions.

\section{Derivative expansion and loop expansion}
\label{sec:setup}

One derivation of the flow equation for the Legendre effective action starts by adding an infrared cutoff $R_\La(q)$ \cite{Wetterich:1992}. Then in perturbation theory, massless propagators with this additive IR (infrared) cutoff appear as 
\be\label{propIR} \Delta_{IR}(q,\La)= \frac{ C_{IR}(q^2/\La^2) }{q^2} = \frac1{q^2+R_\La(q)}\, \ee
where $C_{IR}$ is the corresponding multiplicative IR cutoff, which thus must satisfy
\be \label{CIRnorm} C_{IR}(0)=0\qquad \text{and}\qquad \lim_{u\to\infty}C_{IR}(u)=1\,.\ee
 It will be most convenient for us to rewrite the flow equation for the IR cutoff Legendre effective action $\Gamma[\ph]$ \cite{Wetterich:1992,Morris:1993qb} in terms of this propagator. This results in \cite{Morris:1993qb,Morris:1999ba}:
\be\label{ERG} \partial_\Lambda \Gamma = \frac12\text{tr}\left[ \frac{K_\La}{\Delta_{IR}}\left(1+\Delta_{IR}\Gamma^{(2)}\right)^{-1}\right]\,,
\ee
where  $\Gamma^{(2)}[\ph] = \delta^2\Gamma/\delta\ph\delta\ph$ is the Hessian, and $K_\Lambda = -\partial_\La \Delta_{IR}$.\footnote{$K_\La$ is thus just the $\La$ derivative of the corresponding UV cutoff propagator \cite{Morris:1993qb} although we do not need that identification here.} On the right-hand side, tr is a spacetime trace; in momentum space it is an integral over loop momentum. In this paper we will be interested in solving the flow equation perturbatively in the number of loops:
\be \Gamma = \sum_{\ell=0}^\infty\Gamma_\ell\,,\ee
where $\Gamma_\ell$ is evaluated by performing $\ell$ loop momentum integrals.  If $\hbar$ is made explicit, rather than setting it to $\hbar=1$ as usual, then it appears multiplying the right-hand side of the flow equation \eqref{ERG} and the above expansion is then just a Taylor expansion in powers $\hbar^\ell$.

$\Gamma_0$ coincides with the classical action.
We will be interested exclusively in  massless $\ph^4$ theory, where $\ph$ is a single component real scalar field, for which the classical action has the form:
\be 
\label{class}
\Gamma_0 = \int\!\! d^dx \left\{ \frac12 (\partial_\mu\ph)^2 +\frac{\lambda}{4!}\ph^4\right\}\,.
\ee
From \eqref{ERG},  the flow of the one-loop part is then given by
\be 
\label{ERGone}
\partial_\Lambda \Gamma_1 = \frac12\text{tr}\left[ \frac{K_\La}{\Delta_{IR}}\left(1+\Delta_{IR}\Gamma^{(2)}_0\right)^{-1}\right]\,,
\ee
the flow of the two-loop part by
\be 
\label{ERGtwo}
\partial_\Lambda \Gamma_2 = -\frac12\text{tr}\left[ \frac{K_\La}{\Delta_{IR}}\left(1+\Delta_{IR}\Gamma^{(2)}_0\right)^{-1}\!\!\Delta_{IR}\Gamma^{(2)}_1\left(1+\Delta_{IR}\Gamma^{(2)}_0\right)^{-1}\right]\,,
\ee
and so on. These loop corrections have an infinite expansion in the number of fields. We write their vertices in momentum space as 
\be\label{vertices} \frac{\delta^{2m}\Gamma_\ell}{\delta\ph(p_1)\cdots\delta\ph(p_{2m})}=(2\pi)^{d}\delta(p_1+\cdots+p_{2m})\,\Gamma_\ell(p_1,\cdots,p_{2m})\,.\ee
Here we are recognising that, at least perturbatively, the $\ph\mapsto-\ph$ symmetry of \eqref{class} ensures that vertices have an even number of fields.
In this paper we are also interested in solving for the effective action within a derivative expansion:
\be \Gamma = \sum_{n=0}^\infty \Gamma  |_{\partial^{2n}}\,.\ee
For the $O(\partial^{2n})$ approximation, this corresponds to a truncation in which we keep only the Taylor expansion of the vertices \eqref{vertices} up to $O(p^{2n})$. Provided the IR cutoff is smooth (\ie infinitely differentiable) this kind of approximation always exists. 

\section{Exponential cutoff}
\label{sec:expc}

We start with a cutoff  which allows many of the calculations to proceed analytically, and actually gives the best behaviour for the derivative expansion, namely 
\be\label{Rspecial} R_\La(q) = \frac{q^2}{\text{e}^{\,q^2/\La^2}-1} \,.\ee
It is the popular choice of cutoff profile \cite{Wetterich:1992} apart from the missing wavefunction renormalization factor which, in order to keep the discussion as general as possible, is easier to incorporate in a slightly different way. This is also how this cutoff was used in ref. \cite{Morris:1999ba,Morris:2000hm} to derive results about convergence of the derivative expansion for smooth cutoffs at two loops, results we will review and build on later. Those calculations however concentrated only on the derivative expansions required to compute the two-loop $\beta$ function in four dimensions. They had already appeared in ref. \cite{Papenbrock:1994kf} for that purpose, although there they were not analysed from the point of view of convergence of the derivative expansion. 

For this special choice \eqref{Rspecial} of $R_\La$, we have:
\be\label{IRspecial} C_{IR}(q^2/\La^2) = 1 - \text{e}^{-q^2/\La^2} \,,\ee
and thus 
\be\label{Kexp} K_\La(q)= \frac2{\La^3}\,\text{e}^{-q^2/\La^2}\,.\ee
While this form of cutoff results in both $K_\La(q)$ and $\Delta_{IR}(q)$ 
having a Taylor expansion in $q^2$ with an infinite radius of convergence, due to coefficients $\sim1/n!$, \ie that fall factorially, loop momentum integrals beyond one loop, involve integrating those powers of momentum against $K_\La(q)$, which thus tends to cancel that behaviour. For example, in $d=4$ dimensions we have:\eqrf{27.8,p41}
\be 
\label{momint}
\frac12\int\!\!\frac{d^4q}{(2\pi)^4}\, K_\La(q)\, q^{2n} 
= \frac{(n+1)!}{(4\pi)^2}\,\La^{1+2n}\,,
\ee
Indeed we will see in sec. \ref{sec:anypoint} that from two loops onwards, the derivative expansion either has a finite radius of convergence, or diverges. However we will furnish examples where they can be interpreted as  asymptotic series in the sense discussed in the introduction, in particular in such a way that accurate results can nevertheless be extracted from them.

Note that the resulting IR regulated propagator (\ref{propIR},\ref{IRspecial}) is analytic everywhere in the complex $q^2$ plane except at $q^2=\infty$. Since we Taylor expand over momenta to form the derivative expansion, we can therefore expect that it leads to better convergence than other popular choices of cutoff. This is because, as we discuss in sec. \ref{sec:gen}, these other choices lead to IR regulated propagators that have (complex) singularities at finite distances from $q^2=0$. 

\section{Four dimensions}
\label{sec:exp}

In order for the paper to be self-contained, we start by rederiving the results in ref. \cite{Papenbrock:1994kf,Morris:1999ba} for the two-loop $\beta$ function in $d=4$ spacetime dimensions, computed via derivative expansion, using the special choice of cutoff \eqref{Rspecial}. These results provide useful explicit examples, which we add to by extending the calculation to other operators at two loops, and which we use when verifying results derived using more powerful techniques applicable in general to large $O(\partial^{2n})$ behaviour of higher point vertices and higher loop orders.

To get the two-loop $\beta$ function we need to compute the two-loop four-point vertex at zero momentum. To get this we need first to compute the one-loop effective action up to the six-point vertex.  

\subsection{One-loop two-point, four-point and six-point vertices}
\label{sec:oneloop}

From the one-loop flow equation \eqref{ERGone} and the classical action \eqref{class}, and using \eqref{momint}, we get immediately for the flow of the one-loop two-point vertex: \eqrf{3.6}
\be\label{tadflow} \partial_\La \Gamma_1(p,-p) 
= -\frac{\lambda}{2}\int\!\!\frac{d^4q}{(2\pi)^4}\, K_\La(q) 
= -\frac{\lambda}{(4\pi)^2}\La\,.\ee
(This is the one-loop tadpole correction and is actually $p$-independent.)
Since we want a massless theory after we have integrated out all the modes, we have the renormalization condition $\lim_{\La\to0}\Gamma(p,-p)|_{\partial^0}=0$, or equivalently $\lim_{\La\to0}\Gamma(0,0)=0$. This determines the $\La$-integration constant for the above equation, and thus:
\be\label{onelooptwo} \Gamma_1(p,-p) = -\frac{\lambda}{(4\pi)^2}\frac{\La^2}{2}\,. \ee

The one-loop four-point vertex at zero momentum, or equivalently its $O(\partial^0)$ part, provides a correction to the classical coupling $\lambda$ causing it to run. We define the running coupling $\lambda(\La)$ using the renormalization condition:
\be\label{deflambda} \lambda(\Lambda) = \Gamma(0,0,0,0)\,.\ee
Thus the one-loop flow equation \eqref{ERGone} gives: \eqrf{1.10}
\be\label{betaoneloop}\beta(\lambda)=\La \partial_\La\lambda = -\frac32\lambda^2\La \int\!\!\frac{d^4q}{(2\pi)^4} \frac{\partial_\La C^2_{IR}(q^2/\La^2)}{q^4} = \frac{3\lambda^2}{(4\pi)^2}  \int^\infty_0\!\!\!du\,\frac{d}{du} C^2_{IR}(u) = \frac{3\lambda^2}{(4\pi)^2}\,,\ee
where here and from now on, it should be understood that $\lambda$ depends on the cutoff scale. 
The above result is the one-loop $\beta$ function and, as we have recalled, it is actually universal, independent of the choice of IR cutoff, since all that is actually required to evaluate the momentum integral are the UV and IR cutoff limits \eqref{CIRnorm}.

For the remaining part of the four-point vertex we write
\be\label{four-point} \Gamma(p_1,p_2,p_3,p_4) = \lambda + \gamma(p_1,p_2,p_3,p_4)\,,\ee
where thus $\gamma$ satisfies $\gamma(0,0,0,0)=0$.
At this point we note that the one-loop flow equation can be written as
\be\partial_\La \Gamma_1 = \frac12\, \partial_\La\text{tr} \ln\left(1+\Delta_{IR}\Gamma^{(2)}_0\right) \ee
(where we have dropped a field-independent vacuum energy contribution). 
For the above $\gamma$, and all the higher-point vertices, $2m>4$, we can use the fact that they vanish in the $\La\to\infty$ limit (they are irrelevant operators)  to see that the $\La$-integration constant is zero and thus integrate this immediately to\footnote{The running of $\lambda(\La)$ can be ignored because it is visible only at two loop order in this expression.}
\be\label{Gammaone} \Gamma_1 = \frac12\, \text{tr} \ln\left(1+\Delta_{IR}\Gamma^{(2)}_0\right)\qquad \text{(valid for $\gamma$ or $2m>4$)}\,.\ee
Expanding the log to second order, we have \eqrf{p4 + 6.5}
\be\label{gaminit} \gamma_1(p_1,p_2,p_3,p_4) = -\frac{\lambda^2}{2}\sum_{i=2}^4\int\!\!\frac{d^4q}{(2\pi)^4}\left\{ \Delta_{IR}(q)\Delta_{IR}(q+p_1+p_i)-\Delta_{IR}^2(q)\right\}\,,\ee
where the second term inside the braces comes from subtracting the $O(\partial^0)$ part.

At this point it is useful to note that, for the special choice of cutoff \eqref{IRspecial},
we can write the propagator as a Gaussian by introducing a Schwinger parameter $a$ \cite{Papenbrock:1994kf,Morris:1999ba}: \eqrf{2.1}
\be\label{Schwinger} \Delta_{IR}(q) = \frac1{\La^2}\int^1_0\!\!\!\!da\ \text{e}^{-aq^2/\La^2}\,.\ee
Substituting this into \eqref{gaminit}, with $b$ the Schwinger parameter for $\Delta_{IR}(q+p)$, reduces the loop-integral to one that is purely Gaussian, hence: \eqrf{5.5), (6.1}
\beal 
\int\!\!\frac{d^4q}{(2\pi)^4}\left\{ \Delta_{IR}(q)\Delta_{IR}(q+p)-\Delta_{IR}^2(q)\right\} & =
\frac1{(4\pi)^2}\int^1_0\!\!\!\!da\,db\,\frac1{(a+b)^2}\left( \text{e}^{-\frac{ab}{a+b}p^2}-1\right)\,,\\
&=\frac1{(4\pi)^2}\int^\infty_1\!\!\!\!\!\!dx\,dy\,\frac1{(x+y)^2}\left( \text{e}^{-\frac{p^2}{x+y}}-1\right)\,.
\eeal
In the second line we changed variables to $x=1/a$, $y=1/b$, after which coefficients of the Taylor expansion in $p^2$ are straightforwardly evaluated: \eqrf{p6}
\be 
\label{oneloopfour}
\gamma_1(p_1,p_2,p_3,p_4) = -\frac{\lambda^2}{2(4\pi)^2}\sum_{i=2}^4 \sum_{n=1}^\infty \frac1{(n+1)!\, n}\left(-\frac{(p_1+p_i)^2}{2\La^2}\right)^n\,.
\ee
As with the other results so far, this agrees with ref.  \cite{Papenbrock:1994kf,Morris:1999ba}.

\begin{figure}[ht]
\centering
\includegraphics[scale=0.15]{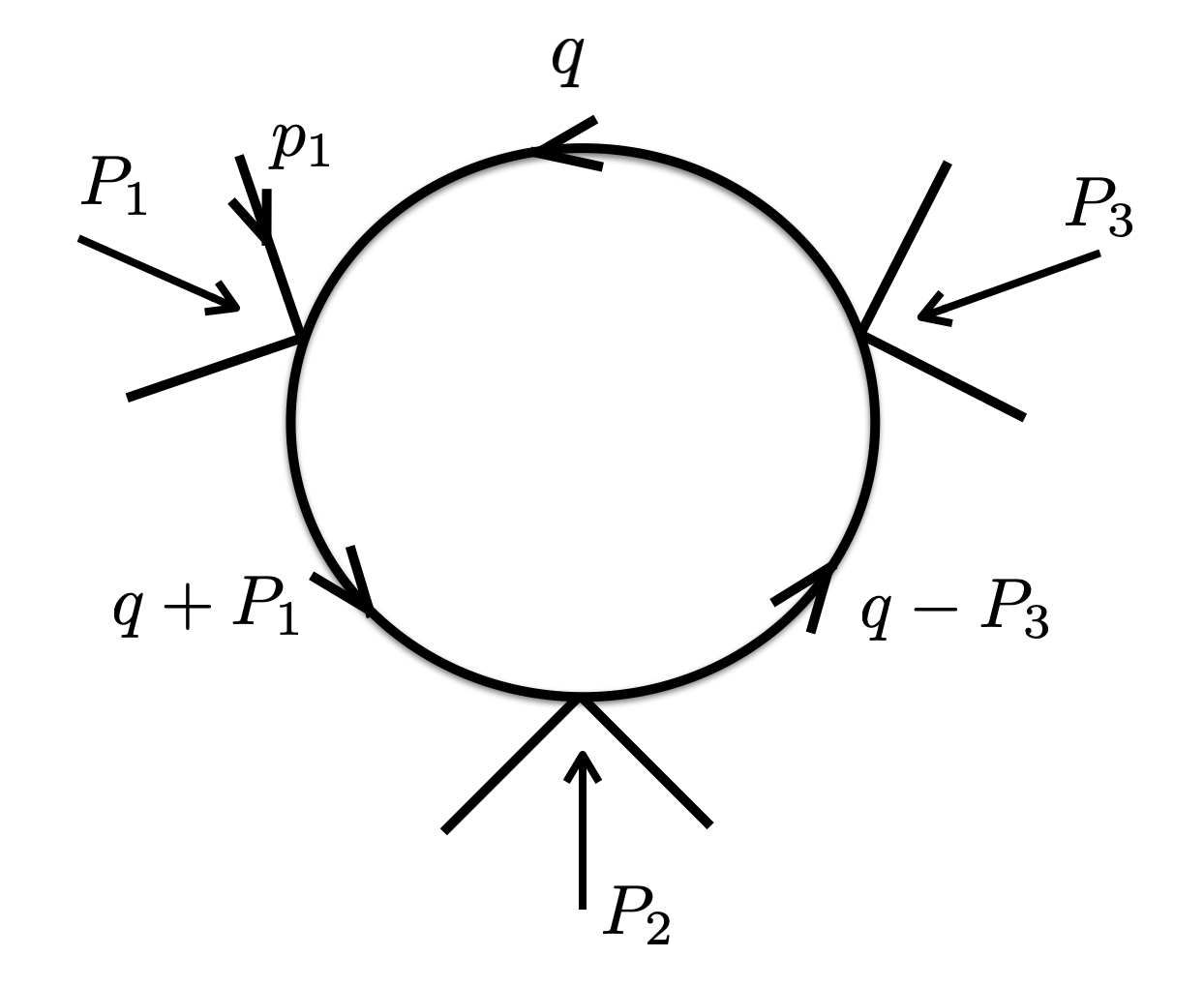}
\caption{The one-loop six-point vertex. By momentum conservation $P_1+P_2+P_3=0$.}
\label{fig:oneloop6pt}
\end{figure}

Finally, we compute the one-loop six-point vertex by expanding \eqref{Gammaone} to third order: \eqrf{7.1}
\be\label{1l6pinit} \Gamma_1(p_1,\cdots,p_6) = \lambda^3 \sum_{\text{pairs}} \int\!\!\frac{d^4q}{(2\pi)^4} \,\Delta_{IR}(q)\Delta_{IR}(q+P_1)\Delta_{IR}(q-P_3)\,,\ee
where the sum is over all 15 ways of arranging the external momenta. 

(Without loss of generality, we can place momentum $p_1$ in the upper-left corner, see fig \ref{fig:oneloop6pt}. Then the complete set of momentum orderings, is given by
summing over pairs of external momenta $P_1=p_{1}+p_{\sigma_2}$, $\{P_2,P_3\}=\{p_{\sigma_3}+p_{\sigma_4}, p_{\sigma_5}+p_{\sigma_6}\}$. Here $\sigma_i$ is a permutation of $i=2,\cdots,6$ subject to pairwise identification inside the $P_2$, and inside $P_3$, and after collecting a factor of 2 by using the reverse loop-momentum routing, which identifies combinations if they are the same under $P_2\leftrightarrow P_3$.) 

Using the Schwinger trick \eqref{Schwinger} (now with parameters $a,b,c$), performing the loop-momentum integral, and Taylor expanding the resulting exponential, we get the derivative expansion form for the one-loop six-point vertex: \eqrf{7.9}
\be\label{oneloopsix}  \Gamma_1(p_1,\cdots,p_6) = \frac{\lambda^3}{(4\pi)^2\La^2}\sum_{n=0}^\infty \frac{(-1)^n}{n!}\sum_{\text{pairs}}
\int^1_0\!\!\!\!da\,db\,dc\,\frac{(abP_1^2+bcP_2^2+caP^2_3)^n}{\La^{2n}(a+b+c)^{n+2}}\,.\ee
This result goes beyond that in ref. \cite{Morris:1999ba} where, in effect, it was computed only for two non-vanishing $p_i$ . However, unlike for $\gamma_1$,  for this general case and at general $O(\partial^{2n})$, it is not possible to evaluate the integrals over Schwinger parameters in closed form.

Note that for these one-loop vertices the derivative expansion always converges for sufficiently small external momenta $p_i$. In fact with this exponential cutoff (\ref{Rspecial},\ref{IRspecial}), the coefficients of the $O(\partial^{2n})$ terms fall factorially with $n$ and thus the radius of convergence is infinite, \ie the derivative expansion converges for all momenta $p_i$.

\subsection{Two-loop self-energy and four-point vertex at zero momentum}
\label{sec:twoloop}

That is no longer true when we go to higher loops. We start by computing the two-loop two-point vertex, using the two-loop flow equation \eqref{ERGtwo}. Substituting the one-loop four-point vertex \eqref{oneloopfour} gives: \eqrf{21'.5}
\beal \partial_\La \Gamma_2(p,-p) &= -\frac12\int\!\!\frac{d^4q}{(2\pi)^4}K_\La(q)\,\gamma_1(p,q,-q,-p)\nn\\ &= \frac{\lambda^2}{(4\pi)^2\La^3}\int\!\!\frac{d^4q}{(2\pi)^4}\text{e}^{-q^2/\La^2}\sum_{n=1}^\infty \frac{(-1)^n}{(n+1)!\, n}\left(\frac{(p+q)^2}{2\La^2}\right)^n\,,
\label{twolooptwo}
\eeal
where we recall that the $n^\text{th}$ term is the $O(\partial^{2n})$ part of the derivative expansion.

Note that the tadpole integral \eqref{tadflow}, formed using the constant part of the four-point vertex \eqref{four-point}, has already been computed:
\be\label{tadpole} \partial_\Lambda \Gamma(p,-p)|_\text{tadpole} = -\frac{\lambda(\La)}{(4\pi)^2}\La\,.\ee
At two loops its $\La$ integral is no longer \eqref{onelooptwo} since the running of the coupling now matters, but the correct expression at this order is straightforward to find:
\beal \Gamma_1(p,-p)+\Gamma_2(p,-p)|_\text{tadpole} &= -\frac1{(4\pi)^2}\int^\La_0\!\! d\La_1\, \La_1\lambda(\La_1)\\  &=  -\frac{\lambda(\La)}{(4\pi)^2}\frac{\La^2}{2}+\frac1{2(4\pi)^2}\int^\La_0\!\! d\La_1\,  \La_1 \beta(\La_1)
=-\frac{\lambda(\La)}{(4\pi)^2}\frac{\La^2}{2}+\frac{3\lambda^2}{(4\pi)^4}\frac{\La^2}{4}\nn \,,\eeal
where the second equality follows by integration by parts, and the third by substituting the one-loop $\beta$ function \eqref{betaoneloop}, recognising that we can discard the higher loop contributions at this order and thus treat the $\lambda^2$ in the last term as independent of $\La$. Iterating in a similar way we can derive the form of this contribution at higher loops also.
However the important point for us here, and later, is that the derivative expansion plays no r\^ole in the tadpole integral over the constant part of the four-point vertex.

To evaluate \eqref{twolooptwo} we use the identity \eqrf{p23}
\be 
\label{qpidentity}
I_n:=\int\!\!\frac{d^4q}{(2\pi)^4}\, (q+p)^{2n}\, \text{e}^{-q^2/\La^2} = 
 \frac{n!(n+1)!}{(4\pi)^2}\La^{4+2n} \sum_{r=0}^n \frac{p^{2r}}{(n-r)!r!(r+1)!\La^{2r}}\,,
\ee
which follows from: \eqrf{p23}
\besp
\label{derivation}
\sum_{n=0}^\infty\frac{I_n}{n!} \frac{\Delta^n}{\La^{2n}} = \int\!\!\frac{d^4q}{(2\pi)^4}\,  \text{e}^{-q^2/\La^2+(p+q)^2\Delta/\La^2} \\
= \frac{\La^4}{(4\pi)^2(1-\Delta)^2}\exp\left(\frac{\Delta}{1-\Delta}\frac{p^2}{\La^2}\right) = 
\frac{\La^4}{(4\pi)^2}\sum_{k=0}^\infty\frac1{k!}\frac{p^{2k}}{\La^{2k}}\frac{\Delta^k}{(1-\Delta)^{k+2}}
\,.
\eesp
Here $\Delta$ is a derivative expansion counting parameter, with $\Delta^n$ giving the $O(\partial^{2n})$ part. Taylor expanding $(1-\Delta)^{-k-2}$ and collecting terms then gives \eqref{qpidentity}.

 Substituting \eqref{qpidentity} into \eqref{twolooptwo} and reordering the summations gives \eqrf{21'.5),(37.10),p48}
 \be 
 \label{twlotw}
\partial_\La \Gamma_2(p,-p) =\frac{\La\lambda^2}{(4\pi)^4}\sum_{r=0}^\infty \frac1{r!(r+1)!}\left(\frac{p^2}{\La^2}\right)^r\!\!\sum_{n=\max(r,1)}^\infty \frac{(-1)^n(n-1)!}{2^n(n-r)!}\,,
\ee
where the outer sum is over the explicit power of $p$ in the Taylor expansion of the two-point vertex, and in the inner sum we have taken care to ensure that $n$ still labels the $O(\partial^{2n})$ part of the derivative expansion. Although the outer sum is thus, in position space, a sum over powers $\partial^{2n}$, the derivative expansion is of course now more than this. The integral over loop momentum $q$ of powers of $q$ coming from the derivative expansion of the one-loop four-point vertex in \eqref{twolooptwo} yields, via the integral  \eqref{momint} against $K_\La$, purely numerical contributions for the coefficient of a given power of $p^2$ in the above vertex, and this gives rise to the inner sum.

The above can now be straightforwardly integrated with respect to $\La$. The integration constants depend on the power of $p$. The $O(p^0)$ term is the two-loop mass term which must vanish at $\La=0$, as we already saw in sec.  \ref{sec:oneloop}. The $O(p^2)$ part integrates to a log, thus requiring the standard arbitrary finite energy scale $\mu$; its coefficient is proportional to the two-loop anomalous dimension and was already computed as a derivative expansion in ref. \cite{Morris:1999ba,Papenbrock:1994kf}, see also app. \ref{app:compold}. The remaining powers all correspond to irrelevant operators which must vanish in the $\La\to\infty$ limit. Thus we find: \eqrf{37.10),p48}
\besp 
\Gamma_2(p,-p) = \frac{\lambda^2}{(4\pi)^4}\Lambda^2\sum_{n=1}^\infty\frac{(-1)^n}{n\,2^{n+1}} +\frac{\lambda^2}{(4\pi)^4}p^2\ln\frac{\La}{\mu}\sum_{n=1}^\infty\frac{(-1)^n}{2^{n+1}}\\ -\frac{\lambda^2}{(4\pi)^4}\La^2\sum_{r=2}^\infty\frac1{(r-1)r!(r+1)!}\left(\frac{p^2}{\La^2}\right)^{\!r}\sum_{n=r}^\infty \frac{(-1)^n(n-1)!}{2^{n+1}(n-r)!}\,.
\label{twlotwint}
\eesp

Apart from the $O(p^2)$ part \cite{Morris:1999ba,Papenbrock:1994kf} these results are new. From the above equation we can already draw some important conclusions. At one loop the derivative expansion is just a Taylor expansion in momenta $p_i$ and thus is a controlled expansion in small parameters namely $p_i/\La$. But from two loops, this is no longer the case: the derivative expansion is a numerical expansion with no control parameter. We can regard each series as an expansion over $\Delta^n$, as above. Then this is a control parameter and in fact all the series in \eqref{twlotwint} have radius of convergence $\Delta=2$. But of course, $\Delta$ is only a bookkeeping parameter here, since in fact $\Delta=1$. However, since $\Delta=1$ is within the radius of convergence, all the series converge (absolutely), although higher derivative operators, $p^{2r}$, converge more slowly for larger $r$ by a factor of $\sim n^r$ at large $n$ (see also app. \ref{app:compold}).  \eqrf{p39}

However, recall from sec. \ref{sec:oneloop} that the one-loop expansions actually had an infinite radius of convergence. Thus the above result shows in fact a dramatic deterioration. The reason for the deterioration is easily seen in the $p^{2r}$ term in \eqref{qpidentity}: at large $O(\partial^{2n})$ its coefficient grows as $(n+1)!\,n^r $. Thus a momentum integral over the cutoff supplies a factorially growing term that cancels the inverse factorial in the one-loop Taylor expansions we saw in sec. \ref{sec:oneloop}, just as we described in general below \eqref{momint}.

On the other hand, after resumming these numerical series we have:
\be 
\label{2l2}
\Gamma_2(p,-p) = \frac{\lambda^2}{(4\pi)^4}\left\{\frac{\Lambda^2}2 \ln\frac23 -\frac{p^2}{6}\ln\frac{\La}{\mu} -\frac{\Lambda^2}2 \sum_{r=2}^\infty \frac1{(r+1)!\, r(r-1)}\left(-\frac{p^2}{3\La^2}\right)^{\!r}\right\}\,,
\ee
and thus the Taylor expansion in external momentum continues to have an infinite radius of convergence. (Recall that the one-loop self-energy was just a tadpole, \eqref{onelooptwo}, so had no external momentum to Taylor expand, but if we compare the above to the four-point vertex at one loop, \eqref{oneloopfour}, we see that the above converges faster.) 

We finish by computing the flow of the two-loop four-point vertex at zero momentum. From the renormalization condition \eqref{deflambda}, this is $\partial_\La\lambda$ at this order, and from the two-loop part \eqref{ERGtwo} of the flow equation, it is given by:
\besp\label{lambdaflowtwo} \La\partial_\La\lambda |_{\ell=2} 
=\\ \La\int\!\!\frac{d^4q}{(2\pi)^4}\, K_\La(q) \left\{ -9\lambda^2\Gamma_1(q,-q) \Delta_{IR}^2(q)+6\lambda\Delta_{IR}(q)\gamma_1(q,-q,0,0)-\frac12\Gamma_1(q,-q,0,0,0,0)\right\}\,.\eesp
The integral over the first term in braces can be written as:
\be\label{lambdatwofirstterm} 3\La\lambda^2\Gamma_1(0,0)\int\!\!\frac{d^4q}{(2\pi)^4}\, \partial_\La\Delta^3_{IR} = \frac{9\lambda^3}{(4\pi)^4}\int^\infty_0 \frac{du}{u}\text{e}^{-u}\left(1-\text{e}^{-u}\right)^2 = 9\frac{\lambda^3}{(4\pi)^4}\ln\frac43\,,\ee
where we used the formula \eqref{onelooptwo}. In the second term in braces in \eqref{lambdaflowtwo}, we recognise that only the momentum-dependent part \eqref{oneloopfour} of the four-point vertex \eqref{four-point} contributes, because the momentum-independent part was already accounted for in the one-loop contribution \eqref{betaoneloop} to the $\beta$ function. Substituting the momentum-dependent part \eqref{oneloopfour}, the result can be evaluated using the identity \eqref{momint}. Alternatively we have directly:
\be\label{la2ter2deriv} -12\frac{\lambda^3}{(4\pi)^4}\int^\infty_0\!\!\!\!du\,\text{e}^{-u}\left(1-\text{e}^{-u}\right)\sum_{n=1}^\infty\frac{1}{(n+1)!n}\left(-\frac{u}{2}\right)^n=
 -12\frac{\lambda^3}{(4\pi)^4}\sum_{n=1}^\infty\frac{(-1)^n}{n(n+1)}\frac1{2^n}\left(1-\frac1{2^{n+1}}\right)\,.\ee
This series converges. Substituting $\Delta^n$ again for bookkeeping purposes, the series has radius of convergence $\Delta=2$, like the two-point vertex \eqref{twlotwint}. It can be summed exactly and the result is:
\be\label{lambdatwosecondterm} 6\frac{\lambda^3}{(4\pi)^4}\left(6\ln3+4\ln2-5\ln5-1\right)\,.\ee
For the final term in braces in \eqref{lambdaflowtwo}, we use the Schwinger integral expression \eqref{oneloopsix} which implies: \eqrf{12.8),p17}
\be\label{1l6ptintegral} \Gamma_1(q,-q,0,0,0,0) = \frac{\lambda^3}{(4\pi)^2\La^2}\left\{ 15\int^1_0\!\frac{da\,db\,dc}{(a+b+c)^2}+12\sum_{n=1}^\infty\frac{(-q^2)^n}{n!\La^{2n}}\int^1_0\!\!\!da\,db\,dc\, \frac{a^n(b+c)^n}{(a+b+c)^{n+2}}\right\}\,.\ee
For the first term in braces above, each integral is straightforward, with result $45\ln\frac43$. In the second term, we substitute $a=1/x$, $b=1/y$, $c=1/z$ so that the integral becomes: \eqrf{p18}
\be\label{intermediates}
\int^\infty_0\frac{dx\,dy\,dz\,(y+z)^n}{(xy+yz+zx)^{n+2}} = \frac1{n+1}\int^\infty_0\frac{dy\,dz\,(y+z)^{n-1}}{(y+z+yz)^{n+1}}
= \frac1{n+1}\int^\infty_1\frac{dz}{z^2}\int^{1+z}_{\frac{1+2z}{1+z}}\frac{du}{u^{n+1}}\,.\ee
In the second step we recognised that the $x$ integral was straightforward.
In the last step we changed variables from $y$ to $u=(y+z+yz)/(y+z)$, after which the $u$ integral is straightforward. Finally changing variables $z=1/a$ transforms the integral to: \eqrf{p18,(18.10)}
\besp 
\frac1{n(n+1)}\int^1_0\!\!da\,\left\{ \left(1-\frac1{2+a}\right)^n-\left(1-\frac1{1+a}\right)^n\right\}
=\\ \frac1{n(n+1)}\left\{ n\ln\frac43+\sum_{r=2}^n\binom{n}{r}\frac{(-1)^r}{r-1}\left(\frac1{2^{r-2}}-\frac1{3^{r-1}}-1\right)\right\}
\,,\label{binomexp}
\eesp
which evaluates to the last line after binomial expansion of the brackets. Thus altogether we find: \eqrf{12.4}
\besp
\label{1l6pt}
\Gamma_1(q,-q,0,0,0,0) =\\ 
\frac{\lambda^3}{(4\pi)^2\La^2}\left( 45 \ln\frac43 +12 \sum_{n=1}^\infty \frac{(-q^2)^n}{(n+1)!\La^{2n}}\left\{\ln\frac43+\frac1n\sum^n_{r=2}\binom{n}{r} \frac{(-1)^r}{r-1}\left[\frac1{2^{r-2}}-\frac1{3^{r-1}}-1\right]\right\}\right)\,,
\eesp
(where the final sum vanishes when $n=1$). Notice that we are left with two summations. The outer sum runs over the derivative expansion orders $O(\partial^{2n})$, whilst the inner sum is there to account for different contributions of the same order.

Substituting this back into the expression \eqref{lambdaflowtwo} for the two-loop flow of $\lambda$ and performing the $q$ integral, \eg by using \eqref{momint}, we get, finally, for the last term in \eqref{lambdaflowtwo}:
\be 
\label{la2terlast}
-\frac{\lambda^3}{(4\pi)^4}\left( 45 \ln\frac43 +12\sum_{n=1}^\infty(-1)^n\left\{\ln\frac43+\frac1n\sum^n_{r=2}\binom{n}{r} \frac{(-1)^r}{r-1}\left[\frac1{2^{r-2}}-\frac1{3^{r-1}}-1\right]\right\}\right)\,.
\ee

\subsection{Two-loop beta function and its rate of convergence}
\label{sec:beta}

We now show that these results give the correct two-loop $\beta$ function, as a check on the calculations so far. To do this, we note that \eqref{la2terlast} can be resummed if we invert the order of the summations. Although the derivative expansion converges in this case (later we will derive the precise rate of convergence), the $\ln\frac43\sum_{n=1}^\infty(-1)^n$ term  clearly does not converge on its own, and (thus) nor on its own does the sum over the remaining terms in braces. To overcome this difficulty, we regularise temporarily by reintroducing the derivative expansion counting parameter $\Delta$.
Then,
\be 
\sum_{n=1}^\infty(-\Delta)^n\left\{\ln\frac43+\frac1n\sum^n_{r=2}\binom{n}{r} \frac{(-1)^r}{r-1}\left[\frac1{2^{r-2}}-\frac1{3^{r-1}}-1\right]\right\}
= -\frac{\Delta}{1+\Delta}\ln\frac43-s_1+4s_{1/2}-3s_{1/3}\,,
\ee
where, inverting the order of summations,
\be 
\label{sa}
s_a = \sum_{r=2}^\infty\frac{(-a)^r}{r-1}\sum_{n=r}^\infty\binom{n}{r}\frac{(-\Delta)^n}{n} =\sum_{r=2}^\infty \frac{y^r}{r(r-1)}=(1-y)\ln(1-y)+y\,,
\ee
where we have set $y=a\Delta/(1+\Delta)$. Finally, setting $\Delta=1$ again, we see that \eqref{la2terlast} sums to
\be 
\label{lastpartresummed}
\frac{\lambda^3}{(4\pi)^4}\left(30\ln5-42\ln2-27\ln3\right)\,.
\ee
Collecting this and the other two results (\ref{lambdatwofirstterm},\ref{lambdatwosecondterm}) for the terms in braces of \eqref{lambdaflowtwo}, all the logs cancel, leaving:
\be 
\La\partial_\La\lambda |_{\ell=2} = -6\frac{\lambda^3}{(4\pi)^4}\,.
\ee
To convert this to the two-loop $\beta$ function, we need to take care of the wavefunction renormalization $\ph\mapsto Z^{1/2}\ph$, where
\be 
Z= 1-\frac{\lambda^2}{(4\pi)^4}\ln\frac\La\mu \sum_{n=1}^\infty\frac{(-1)^n}{2^{n+1}} =
1+\frac{\lambda^2}{6(4\pi)^4}\ln\frac\La\mu\,,
\ee
as required to cancel the $p^2$ term in the two-loop two-point vertex \eqref{twlotwint}, and thus keep the kinetic term normalized. Thus, with our definition \eqref{deflambda} of $\lambda$, and recalling the one-loop result \eqref{betaoneloop}, we get
\be 
\beta = \La\partial_\La (\lambda Z^2) = \frac{3\lambda^2}{(4\pi)^2} +\left(\frac13-6\right)\frac{\lambda^3}{(4\pi)^4} = \frac{3\lambda^2}{(4\pi)^2} -\frac{17}{3}\frac{\lambda^3}{(4\pi)^4}\,,
\ee
the well-known universal result for the $\beta$ function to two loops (see \eg ref. 
\cite{ZinnJustin:2002ru}).

In app. \ref{app:compold}, we compare these results to the way they were derived in ref. \cite{Morris:1999ba}, see also \cite{Papenbrock:1994kf}. Here, we comment on convergence of these derivative expansions. Reinstating the derivative expansion counting parameter $\Delta$, we saw below \eqref{twlotwint} that the radius of convergence for the two-loop two-point vertex was $\Delta=2$. That is also true of the contribution \eqref{la2ter2deriv} to the two-loop $\beta$ function, equivalently the (flow of the) two-loop four-point vertex at zero momentum. However the final contribution \eqref{la2terlast} controls the overall convergence rate because it turns out to have the smallest radius of convergence. In ref. \cite{Morris:1999ba} it was reported without proof that the $n^\text{th}$ term in this contribution fell faster than $(2/3)^n/n$ (and thus has at least a radius of convergence of $\Delta=3/2$). We will now confirm this, and in fact determine the exact asymptotic fall-off for the zero-momentum two-loop four-point vertex. This also introduces the strategy we will be able to use more generally. To determine the asymptotic behaviour at large order in the derivative expansion, we return first to the one-loop six-point vertex in its special momentum configuration \eqref{1l6pt}. This already contains the expression for the $n^\text{th}$ term in the series in \eqref{la2terlast}, whose behaviour we are aiming to determine asymptotically. The coefficient arose in particular from the second Schwinger-parameter integral in \eqref{1l6ptintegral}:
\beal 
\int^1_0\!\!\!da\,db\,dc\, \frac{a^n(b+c)^n}{(a+b+c)^{n+2}} &= \int^1_0\!\frac{da\,db\,dc}{a^2(b+c)^2} \left(\frac1a+\frac1{b+c}\right)^{-n-2}\nn\\ 
&= \frac14\left(\frac23\right)^{n+2}\int_0\!\!d\alpha\,d\beta\,d\gamma\, \left(1+\frac{2\alpha}{3}+\frac{\beta}{6}+\frac{\gamma}{6}\right)^{-n-2}+\cdots\nn\\
&= \frac6{n^3}\left(\frac23\right)^{n}\!\!+\cdots\,.\label{largenint}
\,,
\eeal
Writing the integral as the second expression in the top line, 
makes it clear that at large $n$, the integral is dominated by the corner $a\to1$, $b\to1$ and $c\to1$, where $1/a$ and $1/(b+c)$ are smallest. Writing $a=1-\alpha$, $b=1-\beta$, 
and $c=1-\gamma$, to get the leading behaviour in $n$, we need only keep first order in $\alpha, \beta, \gamma$ inside the $(n+2)^\text{th}$ power and zeroth order elsewhere. This gives us the second line, where  the ellipses stand for terms that are subleading in the large $n$ limit. Finally, we recognise that after integration only the bottom boundary of the integral makes a contribution at leading order. This makes all three integrals straightforward, giving the final line. 
Substituting into \eqref{1l6ptintegral} we get immediately the large $O(\partial^{2n})$ behaviour for the special momentum configuration one-loop six-point vertex:
\be\label{largen6pt} \Gamma_1(q,-q,0,0,0,0)|_{\partial^{2n}} = \frac{\lambda^3}{(4\pi)^2\La^2}\frac{72}{n!\,n^3}\left(-\frac23\frac{q^2}{\La^2}\right)^n+\cdots\,,
\ee
whilst from the integral's exact expression \eqref{binomexp}, we see that the $O(\partial^{2n})$ term in the two-loop zero-momentum four-point vertex \eqref{la2terlast} behaves as 
\be\label{largen4pt}
\La\partial_\La\Gamma_2(0,0,0,0)|_{\partial^{2n}}=-\frac{\lambda^3}{(4\pi)^4} \frac{72}{n^2} \left(-\frac23\right)^{n} +\cdots\,,\ee
where we recognise that this $O(\partial^{2n})$ contribution dominates at large $n$ since it has radius of convergence $\Delta=3/2$, whereas the other contributions to this four-point vertex have $\Delta=2$. We see that, as stated in ref. \cite{Morris:1999ba}, this contribution to the two-loop vertex indeed falls faster than $(2/3)^n/n$. We
have confirmed \eqref{largen4pt}, by checking numerically that its ratio with the exact $O(\partial^{2n})$ term, the sum of the $n^\text{th}$ terms in \eqref{la2ter2deriv} and \eqref{la2terlast}, tends to one at large $n$.

\section{Any-point vertices in any dimension}
\label{sec:anypoint}

The derivative expansions we have investigated so far, are all convergent. However in these results are hints of where to look to find expansions that are not so well behaved. 

Firstly the expansions found so far at two loops, do not all converge at the same rate. In fact it is this final one, \eqref{largen4pt}, that has the slowest convergence. We will shortly provide an explanation for this, which will then motivate our choice of an infinite class of other operators to analyse. At the same time, we take the opportunity to generalise the analysis away from four spacetime dimensions\footnote{or space dimensions in the context of statistical physics} to general dimension $d$.

Without the cutoff regularisation, Taylor expansions in external momenta $p$ do not exist (for any non-trivial vertex) because, taken to sufficiently high order, their coefficients are IR divergent loop integrals. In the one-loop vertices these loop integrals are simply:
\be 
\label{IRint}
\int\!\! \frac{d^dq}{(2\pi)^d}\, \frac1{q^{2r}}\,,
\ee
which are IR divergent when $r\ge d/2$. In the previous section, we found that the $O(\partial^{2n})$ contribution \eqref{la2terlast} to the last term in  \eqref{lambdaflowtwo}, the two-loop flow of $\lambda$, provides the slowest convergence, namely
\eqref{largen4pt}, amongst those we have studied so far. This behaviour \eqref{la2terlast} is directly inherited from the one-loop six-point vertex \eqref{1l6pt}. With the above in mind, it is easy to appreciate why, amongst those one-loop vertices we have considered so far, the one-loop six-point vertex provided us with the slowest converging  expansion. It is simply that without the cutoff, its coefficients are the most IR divergent.\footnote{\textit{E.g.} even the $p^0$ term, gives $r=3$ in \eqref{IRint}, and thus diverges quadratically in $d=4$ dimensions, whilst for the four-point  vertex this is a logarithmic divergence, and the two-point vertex is IR divergent only starting at $O(p^2)$.} Regularisation with the cutoff gives us a finite result, but the more divergent the integral without it, the larger that result will be when the cutoff is introduced.

Since the power in \eqref{IRint} is simply given by dimensions as $2r=d_{\cal O}-d$, where $d_{\cal O}$ is the dimension of the operator, we expect that the more irrelevant the operator the poorer the convergence of its derivative expansion at one loop. In turn these will lead to yet poorer convergence at higher loops. 

\subsection{One loop}
\label{sec:oneloops}

This motivates studying the derivative expansion of a general $2m$-point vertex $\Gamma_1(p,-p,0^{2m-2})$  (here $0^{2m-2}$ stands for $2m-2$ vanishing momentum arguments). As in sec. \ref{sec:twoloop} we can then calculate contributions at two loops for $(2m-2)$-point vertices at zero momentum, by sewing together the momentum-dependent legs. Furthermore, mimicking the calculation \eqref{twolooptwo} of the two-loop self-energy, we can generalise this to compute some contributions at non-zero external momenta. In this way we will also be able to test how well behaved the expansion is for derivative operators.

\begin{figure}[ht]
\centering
\includegraphics[scale=0.1]{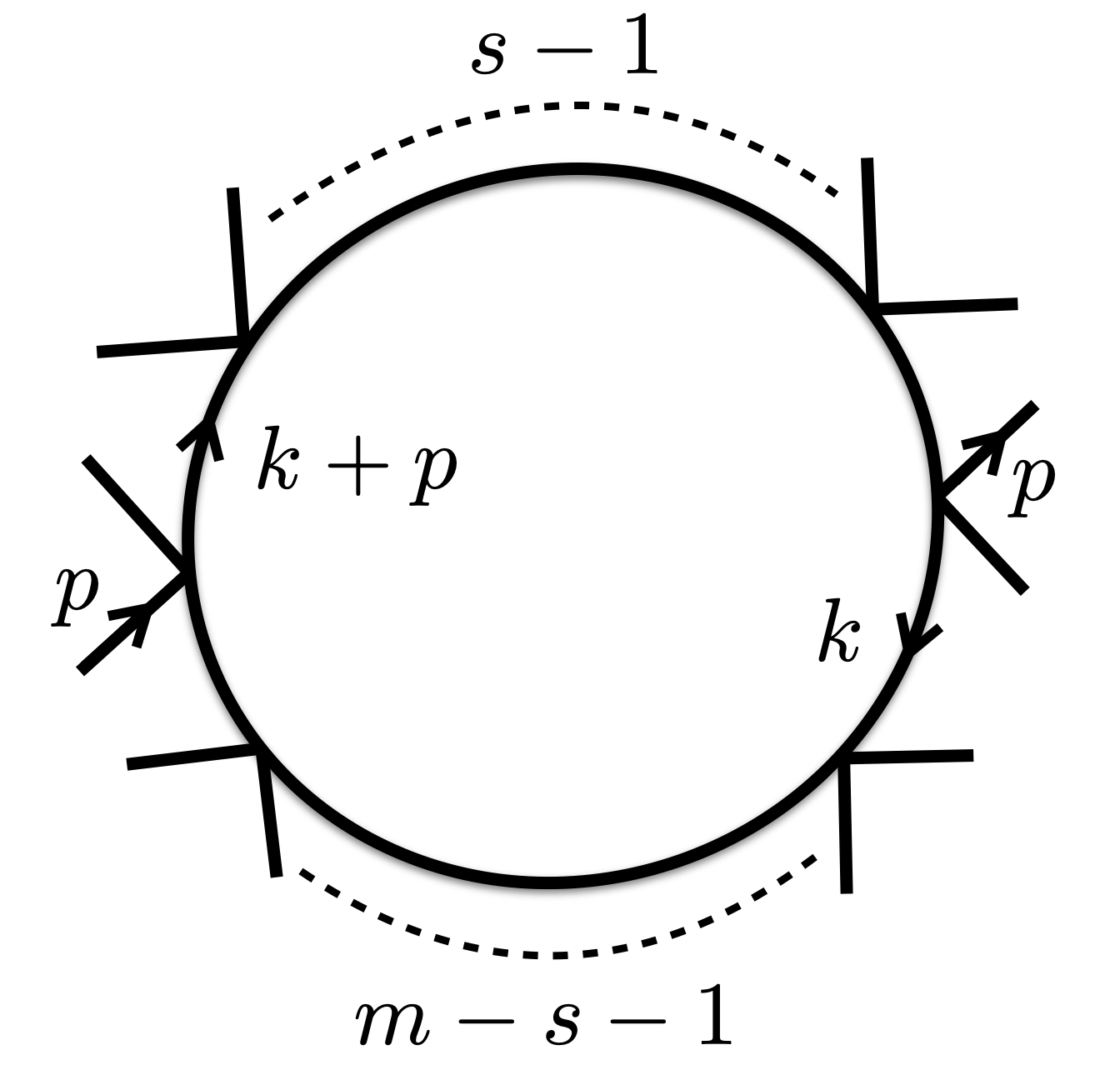}
\caption{One-loop Feynman diagrams that contribute to $\Gamma_1(p,-p,0^{2m-2})$. The unlabelled external legs carry zero momentum. This includes the $s-1$ $\lambda\ph^4$-vertices around the top of the loop, and the $m-s-1$ $\lambda\ph^4$-vertices around the bottom of the loop.}
\label{fig:oneloop2m}
\end{figure}

Expanding the one-loop action \eqref{Gammaone} to $m^\text{th}$ order, we get
\be\label{oneloop2m} \Gamma_1(p,-p,0^{2m-2}) = -(2m-2)!\left(-\frac\lambda2\right)^{m}\!\!\int\!\! \frac{d^dk}{(2\pi)^d}\,\Big\{ \Delta_{IR}^m(k)+2\,\Pi_m(k,p)\Big\}\,,\ee
where 
\bea
\label{Pi}
\Pi_m(k,p) = &2\sum_{s=1}^{\frac{m-1}{2}}\Delta_{IR}^{m-s}(k)\Delta_{IR}^s(k+p)\,,\phantom{+\Delta_{IR}^{\frac{m}{2}}(k)\Delta_{IR}^{\frac{m}{2}}(k+p)}\  &(m\ \text{odd})\,,\nn\\
= &2\sum_{s=1}^{\frac{m}{2}-1}\Delta_{IR}^{m-s}(k)\Delta_{IR}^s(k+p)+\Delta_{IR}^{\frac{m}{2}}(k)\Delta_{IR}^{\frac{m}{2}}(k+p)\,, &(m\ \text{even})\,.
 \eea
The $\Delta^m_{IR}(k)$ term in \eqref{oneloop2m} corresponds to the case where the external momentum $p$ enters and leaves the same $\lambda\ph^4$-vertex. The other contributions are illustrated in fig. \ref{fig:oneloop2m} and listed in eqn. \eqref{Pi}; the factor of 2 in front of the sums coming from combining terms related by reversing the loop-momentum routing ($k\mapsto -k-p$).

Note that the different contributions that sum up to give $\Gamma_1(p,-p,0^{2m-2})$ are thus effectively labelled by the power $s$ of propagators carrying momentum $k+p$. After using the Schwinger parametrisation \eqref{Schwinger}, their loop-momentum integrals are purely Gaussian and thus one readily finds:
\be 
\label{intprops}
\int\!\! \frac{d^dk}{(2\pi)^d}\,\Delta_{IR}^{m-s}(k)\Delta_{IR}^s(k+p) = \frac{\La^{d-2m}}{(4\pi)^{d/2}}\sum_{n=0}^\infty\frac{I^n_{m,s}}{n!}\left(-\frac{p^2}{\La^2}\right)^n\,,
\ee
where the Schwinger integral is
\be 
\label{Inms}
I^n_{m,s} = \int^1_0\frac{d^{\,m-s\!}a\,d^sb}{A^{d/2}B^{d/2}}\left(\frac1A+\frac1B\right)^{-n-d/2}\,,
\qquad A=\sum_{i=1}^{m-s}a_i\,,\quad B=\sum_{j=1}^sb_j\,,\ee
and thus for $n> 0$ we have the exact expression:
\be
\label{1l2mex}
\Gamma_1(p,-p,0^{2m-2})|_{\partial^{2n}} =
 -4(2m-2)!\frac{\La^{d-2m}}{(4\pi)^{d/2}}\left(-\frac{\lambda}{2}\right)^m \frac{J^n_m}{n!}\left(-\frac{p^2}{\La^2}\right)^{\!n}\,,
\ee 
where the Schwinger integral contribution is summarised in:
\be
J^n_m =  \frac12 I^n_{\frac{m}{2},\frac{m}{2}}+\sum^{\frac{m}{2}-1}_{s=1} I^n_{m,s}\,,\quad\ (m\ \text{even})\,;\qquad
J^n_m= \sum^{\frac{m-1}{2}}_{s=1} I^n_{m,s}\,,\quad (m\ \text{odd})\,,
\ee
As in \eqref{largenint}, we have written \eqref{Inms} in a form where it is clear that at large $O(\partial^{2n})$, keeping finite the other parameters ($d$ and $m$), the Schwinger integral is dominated by the corner where the parameters are all close to their upper limit. Thus, writing $a_i=1-\alpha_i$ and $b_j=1-\beta_j$, we get the leading behaviour in $n$ by keeping first order in $\alpha_i$ and $\beta_j$ inside the large power, and zeroth order elsewhere. Then the integrals are straightforward because, as before, only the bottom boundary makes a contribution to leading order. Hence we find: \eqrf{p60,62}
\beal
I^n_{m,s}
&= \frac{s^n(m-s)^n}{m^{n+d/2}}\int_0\!\!d^{\,m-s}\!\alpha\,d^s\!\beta\left(1+\frac{s}{m(m-s)}\sum^{m-s}_{i=1}\alpha_i+\frac{m-s}{ms}\sum_{j=1}^s\beta_j\right)^{-n-d/2}\!\!\!\!+\cdots\,,\nn\\
&= m^{m-d/2}\left(\frac{m}{s}-1\right)^{m-2s}\frac1{n^m}\left(\frac{s(m-s)}{m}\right)^{\!n}+\cdots\,,
\label{asym}
\eeal
where, once again, the ellipses stand for terms that are subleading in the large $n$ limit. 

Since $n\gg m$ is not explicitly used in getting this result, one may wonder whether this limit is actually needed. To see why it is needed, note that for example in the integral above, the upper boundary cannot be neglected if $s\sim m\sim n$, and furthermore the $m$ Schwinger integrals raise the power $-n-d/2$ to  $-n+m-d/2$. Finally, it can also be shown that the higher orders of $\alpha_i$ and $\beta_j$ that we have dropped, cannot be neglected if $s\sim m\sim n$.

Notice that at leading order, the large $n$ behaviour of \eqref{asym} does not depend on the 
spacetime dimension $d$; it only enters in the prefactor. Secondly, notice that the leading behaviour at large $n$ comes from the contribution whose power $s$ is closest to $m/2$,\footnote{This follows for example by noting that $s(m-s)= m^2/4-(s-m/2)^2$.} namely $s=(m-1)/2$ or $m/2$  for $m$ odd or even respectively. The reason for this is already clear from the left hand side of \eqref{intprops}: the more propagators that carry the external momentum $p$, the larger the coefficients in its Taylor expansion. But $s$ cannot be larger than $m/2$, since if it is, reversing the loop-momentum routing maps this back to $s\le m/2$ again.
Thus for \eqref{1l2mex} we find that the leading asymptotic behaviour is given by:
\besp
\label{largen2m}
\Gamma_1(p,-p,0^{2m-2})|_{\partial^{2n}}   = -2(2m-2)!\frac{\La^{d-2m}}{(4\pi m)^{d/2}}\left(-\frac{m}{2}\lambda\right)^m\frac1{n!\,n^m}\left(-\frac{m}4 \frac{p^2}{\La^2}\right)^n+\cdots\qquad(m\ \text{even})\,,\\
 = -4\frac{m+1}{m-1}(2m-2)!\frac{\La^{d-2m}}{(4\pi m)^{d/2}}\left(-\frac{m}{2}\lambda\right)^m\frac1{n!\,n^m}\left(-\frac{m^2-1}{4m} \frac{p^2}{\La^2}\right)^n+\cdots\qquad(m\ \text{odd})\,.
\eesp
The above large $n$ behaviour agrees with the two cases we have already derived in $d=4$ dimensions, namely the case  $m=2$ which agrees with $\gamma_1(p,-p,0,0)$ in \eqref{oneloopfour} (noting that $(n+1)!$ can be replaced by $n!\,n$ to leading order at large $n$), and the case $m=3$ with \eqref{largen6pt}. 

As before, we see that the one-loop derivative expansions have an infinite radius of convergence, thanks to the $1/n!$ factor, although this convergence is slower for larger $m$ -- as expected by the arguments at the beginning of this section. 

\subsection{Two loops}
\label{sec:twoloops}

Analogous to the last contribution in \eqref{lambdaflowtwo} we can now compute a contribution to the zero-momentum two-loop $(2m-2)$-point vertex $\Gamma_2(0^{2m-2})$ via
\be 
\label{qint0}
\partial_\La\Gamma_2(0^{2m-2}) \ni -\frac12 \int \frac{d^dq}{(2\pi)^d} K_\La(q)\, \Gamma_1(q,-q,0^{2m-2})\,.
\ee
Here we write ``$\ni$'' to indicate that this is one of a number of contributions, the others being generalisations of the contributions in \eqref{lambdaflowtwo}, although
 we expect this to be the dominant contribution at large $O(\partial^{2n})$, just as we found in sec. \ref{sec:beta}. However, instead of computing \eqref{qint0} directly, we will make a further generalisation which allows us to use the above results to get at one $O(\partial^{2n})$ contribution to $\Gamma_2(p,-p,0^{2m-4})$ via: 
\be 
\label{qintp}
\partial_\La\Gamma_2(p,-p,0^{2m-4}) \ni -\frac12 \int \frac{d^dq}{(2\pi)^d} K_\La(q)\, \Gamma_1(q,-q,p,-p,0^{2m-4})\,.
\ee
Taylor expanding up to $p^{2r}$ will then furnish examples of two-loop contributions to derivative operators containing $2m-2$ fields and $2r$ derivatives.

\begin{figure}[ht]
\centering
\includegraphics[scale=0.15]{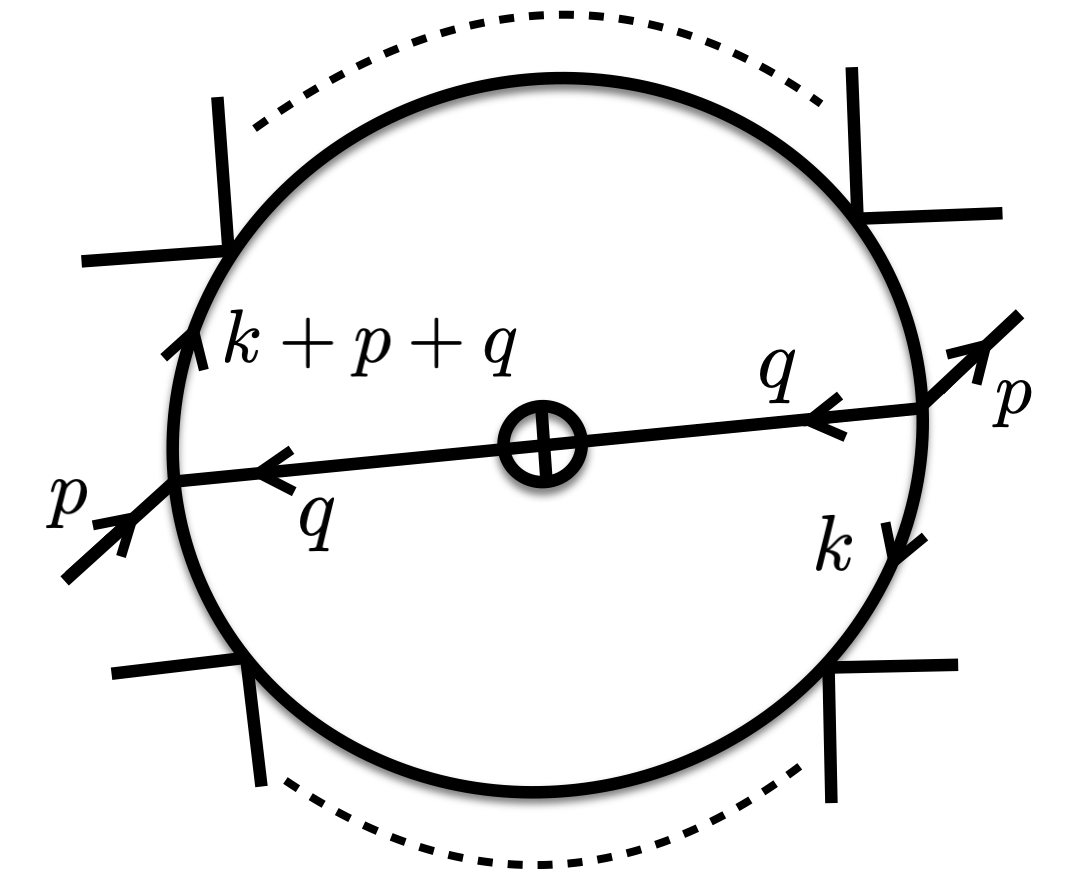}
\caption{A two-loop Feynman diagram contribution to the flow of the $(2m-2)$-point vertex $\Gamma_2(p,-p,0^{2m-4})$, formed by tying together the other external legs next to the $p$-dependent ones in the one-loop $2m$-point vertex of fig. \ref{fig:oneloop2m}, and integrating them over $K_\La(q)$ (indicated here by the crossed circle).  The unlabelled external legs carry zero momentum.}
\label{fig:twoloop2m-2}
\end{figure}

We will only study a part of contribution \eqref{qintp}, namely the part obtained by taking the one-loop $2m$-point vertex $\Gamma_1(p,-p,0^{2m-2})$ and replacing  $p^\mu\mapsto p^\mu+q^\mu$ in \eqref{largen2m}. This corresponds to the contribution where the momentum $p$ and $q$ enter the same $\lambda\ph^4$ vertex and leave together via another such $\lambda\ph^4$ vertex, \ie  whose Feynman diagram is as illustrated in fig. \ref{fig:twoloop2m-2}. To compute it using \eqref{largen2m}, we only need the $d$-dimensional generalisation of the identity \eqref{qpidentity}: \eqrf{64.1}
\be 
\label{qpdidentity}
\int\!\!\frac{d^dq}{(2\pi)^d}\, (q+p)^{2n}\, \text{e}^{-q^2/\La^2} = \frac{\La^{d+2n}}{(4\pi)^{d/2}}\sum_{r=0}^n \frac{\Gamma(n+d/2)}{\Gamma(r+d/2)}\binom{n}{r}\left(\frac{p^2}{\La^2}\right)^{r}\,,
 \ee
which can be derived following the same steps as in \eqref{derivation}. Thus from the exact one-loop result \eqref{1l2mex} (and relabelling $m\mapsto m+1$) we find a sample series of two-loop $O(\partial^{2n})$ contributions to the flow of the $2m$-point vertex:
\be
\label{2l2mex}
\partial_\La\Gamma_2(p,-p,0^{2m-2})|_{\partial^{2n}} \ni
 4(2m)!\frac{\La^{2d-2m-5}}{(4\pi)^{d}}\left(-\frac{\lambda}{2}\right)^{m+1}\!\!\!\! J^n_{m+1}(-1)^n
\sum_{r=0}^n \frac{\Gamma(n+d/2)}{r!\,(n-r)!\,\Gamma(r+d/2)}\left(\frac{p^2}{\La^2}\right)^{\!r}\,,
\ee 
Using the asymptotic result \eqref{largen2m} for the one-loop vertex, we see that at high order $O(\partial^{2n})$, this simplifies to: 
\be
\label{twoloop2m}
\partial_\La\Gamma_2(p,-p,0^{2m-2})|_{\partial^{2n}} \ni C_m\La^{2d-2m-5}\sum_{r=0} \frac{n^{r-m+d/2-2}(-c_m)^n }{r!\,\Gamma\!\left(r+\frac{d}{2}\right)}\,\left(\frac{p^2}{\La^2}\right)^r
+\cdots\,.
\ee
where additionally we have simplified by using $n\gg r$.
In the above,
\be\label{cm} c_m\ =\ \frac{m+1}4\quad(m\ \text{odd})\,,\qquad c_m=\frac{m+1}{4}\left(1-\frac1{(m+1)^2}\right) = \frac{m(m+2)}{4(m+1)}\quad(m\ \text{even})\,,
\ee
and we have collected some prefactors into the coefficient
\beal 
C_{m} &= \frac{(2m)!(m+1)^{m+1}\left(-\lambda\right)^{m+1}}{2^{m}(m+1)^{d/2}(4\pi)^d}\qquad\qquad\quad (m\ \text{odd})\,,\nn\\
 &= \frac{2(m+2)}{m}\,\frac{(2m)!(m+1)^{m+1}\left(-\lambda\right)^{m+1}}{2^{m}(m+1)^{d/2}(4\pi)^d}\quad(m\ \text{even})\,,
 \label{Cmr}
\eeal
so as not to distract from the important point here: the large $n$ dependence in \eqref{twoloop2m}. However we note that the two cases we already derived in $d=4$ dimensions, agree exactly with the above in the large $n$ limit, namely $m=1$ (any $r$), and $(m,r)=(2,0)$. These cases correspond respectively to the flow of the two-point vertex \eqref{twlotw} expanded in powers $p^{2r}$, and the zero-momentum four-point vertex \eqref{largen4pt}.

Integrating \eqref{twoloop2m} so as to study the vertices rather than their flow does not substantially change the analysis. We have instead:
\be\label{2loop2mi} \Gamma_2(p,-p,0^{2m-2})|_{\partial^{2n}} \ni-\frac12 C_m\La^{2d-2m-4}\sum_{r=0} \frac{n^{r-m+d/2-2}(-c_m)^n }{r!\,(m+r+2-d)\Gamma\!\left(r+\frac{d}{2}\right)}\,\left(\frac{p^2}{\La^2}\right)^r
+\cdots\,,\ee
except when $m\le d-2$ in which case we must exclude $r=d-m-2$ from the sum and add
\be 
\label{subst}
C_m \ln\frac{\La}{\mu}\, \frac{n^{3d/2-2m-4}(-c_m)^n}{(d-m-2)!\,\Gamma\!\left(\frac{3d}{2}-m-2\right)}{\left(p^2\right)}^{d-m-2}+\cdots\,.\ee
Similarly, integrating the exact result \eqref{2l2mex} with respect to $\La$ gives:
\besp
\label{2l2mexi}
\Gamma_2(p,-p,0^{2m-2})|_{\partial^{2n}} \ni \\
 -2(2m)!\frac{\La^{2d-2m-4}}{(4\pi)^{d}}\left(-\frac{\lambda}{2}\right)^{m+1}\!\!\!\! J^n_{m+1}(-1)^n
\sum_{r=0}^n \frac{\Gamma(n+d/2)}{(m+r+2-d)r!\,(n-r)!\,\Gamma(r+d/2)}\left(\frac{p^2}{\La^2}\right)^{\!r}\,,
\eesp
except when $m\le d-2$ in which case we must exclude $r=d-m-2$ from the sum and add
\be 
\label{substex}
\frac{4(2m)!}{(4\pi)^{d}}\ln\frac{\La}{\mu}\left(-\frac{\lambda}{2}\right)^{m+1}\!\!\!\! J^n_{m+1}
 \frac{(-1)^n\,\Gamma(n+d/2)}{(d-m-2)!\,(n+m+2-d)!\,\Gamma(3d/2-m-2)}\left(p^2\right)^{d-m-2}.
\ee

\begin{figure}[ht]
\centering
\includegraphics[scale=0.2]{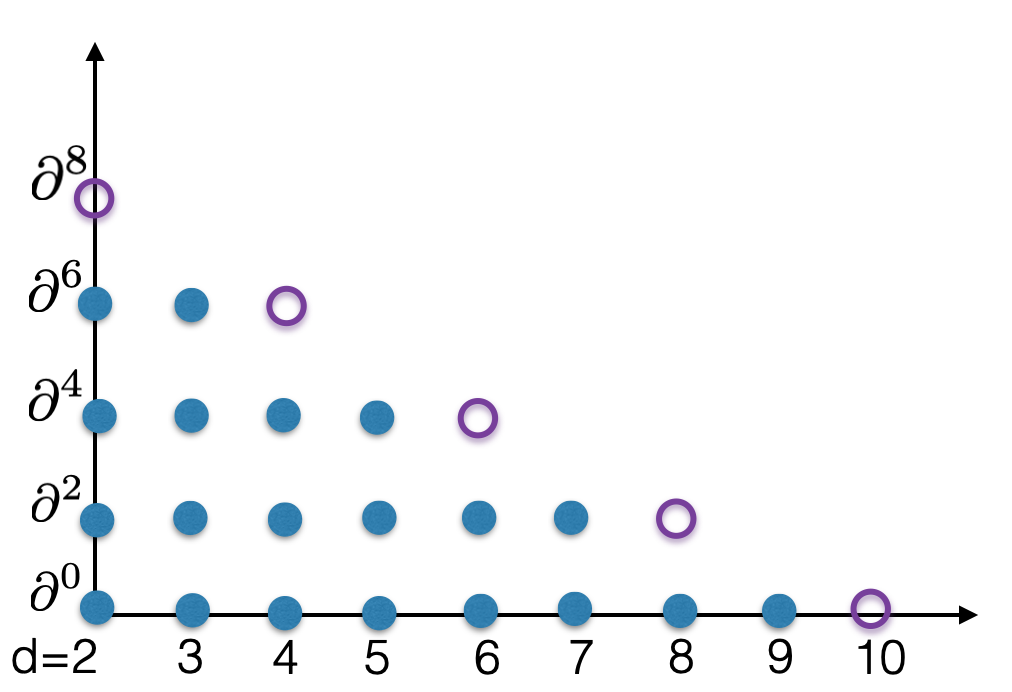}
\caption{The filled circles give the dimensions $d$ in which six-point $2r$-derivative operators have convergent derivative expansions at two loops. The line with no derivatives, \viz $\partial^0$, thus corresponds to the $\ph^6$ operator. The open circles mark special cases where the derivative expansion series is purely oscillatory at leading order in $n$ and thus no longer converges. For all other values of $d$ and $r$, the derivative expansion diverges.
Since the lower-point vertices always have convergent derivative expansions at two loops and at leading order the higher-point vertices have asymptotic expansions  from which accurate results can nevertheless be extracted, the filled circles also mark out the region where, according to the leading order approximation, accurate results can be extracted from the derivative expansion as a whole at two loops.}
\label{fig:derivs}
\end{figure}

Turning now to their $O(\partial^{2n})$ convergence properties, we see that \eqref{2loop2mi} again displays the $n^r$ weakening of convergence in higher derivative operators with $2r$ derivatives, as already commented on below \eqref{twlotwint} and generally in app. \ref{app:compold}. We also see that larger $d$ weakens convergence. However both $r$ and $d$ only affect the power of $n$, and this is subleading compared to exponential dependence on $n$ provided by $(-c_m)^n$, so long as $c_m\ne1$. Thus the value of $c_m$ is crucial.
Since $c_1<1$ and $c_2<1$, it tells us that the derivative expansion converges for the two-point vertices and four-point vertices, containing any number of derivatives $\partial^{2r}$ and in any dimension $d$. On the other hand, for any $(2m\geq 8)$-point vertex, we have that $c_m>1$ and thus their derivative expansions do not converge, for any $r$, or any dimension $d$. 
Separating these two regions is the marginal case of six-point vertices ($m=3$), for which $c_3=1$. In this case convergence for a $\partial^{2r}$ derivative operator requires $d<10-2r$, so that the terms in the oscillating series \eqref{2loop2mi} get smaller with increasing $n$. That leaves the examples displayed as filled circles in fig. \ref{fig:derivs}. If this inequality is not satisfied the series does not converge. This does not necessarily mean that accurate results cannot be extracted, but to find out whether that is possible or not would require going beyond the leading terms; see the discussion in sec. \ref{sec:eightpoint} and the conclusions, sec. \ref{sec:conclusions}.

Although the $(2m\geq 8)$-point vertices have divergent derivative expansions, 
this is not evident in higher order non-perturbative calculations \cite{Balog:2019rrg}, so we already know that their divergent nature must only become visible at even higher order. 
Actually, this feature is already apparent in the leading terms \eqref{twoloop2m}, as we demonstrate more fully in sec. \ref{sec:eightpoint}. It arises because the power of $1/n$ in \eqref{twoloop2m} initially wins out and increasingly so with increasing $m$, such that the $O(\partial^{2n})$ contributions continue to decrease with increasing $n$ up to some high order critical value $n=n_{cr}(m)$ (which is an increasing function of $m$) and only with the next term, \ie once $n=n_{cr}(m)+1$, do they stop decreasing and instead start to diverge (exponentially fast) with $n$. This is the regime where, despite the fact that the derivative expansion ultimately diverges, accurate results can nevertheless be extracted from it because, as we will show, the derivative expansion contributions then form an asymptotic series, in the sense defined in the Introduction, sec. \ref{sec:intro}.  Indeed we then expect that the following well-known rule of thumb \cite{OlverAsymptotics,BerryHowls1990} should apply: that the error can be estimated by the next term in the series, the most accurate result being obtained by truncating the series just short of the smallest term (\ie at $n=n_{cr}(m)-1$). In fact in sec. \ref{sec:asymptotics}, we prove that this rule of thumb works for the leading $O(\partial^{2n})$  behaviour \eqref{2loop2mi} considered as a series on its own.

We will develop the consequences in sec. \ref{sec:eightpoint}. For the moment we note that, because $n_{cr}(m)$ increases with $m$, the limit to the range where accurate results can still be extracted, is determined by the lowest value of $m$ where the derivative expansion does not automatically converge, which is the $m=3$ case again. Here what is required is that the power of $1/n$ remains positive. Thus the boundary is determined as $d<10-2r$ again, just as for convergence of the derivative expansion in six-point vertices. The special cases precisely at the boundary $d=10-2r$, correspond to cases where higher-point vertices have useful asymptotic series, \ie still have derivative expansions that are convergent up to some critical $n=n_{cr}(m)$, but the six-point vertex now has an oscillatory derivative expansion that no longer converges even temporarily. This is so because the only $n$ dependence left in the leading contribution \eqref{2loop2mi}, is the $(-1)^n$ factor. These marginal cases are marked by open circles in fig. \ref{fig:derivs}.

There is one important proviso to all this. We can of course trust the leading terms, only so long as the subleading terms remain much smaller than them. This requires in particular that
$n\gg d,m,r$. Clearly, the leading terms are sufficient to prove properties of the $O(\partial^{2n})$ expansion in the limit $n\to\infty$.
But it turns out that the leading terms alone are not sufficient for estimating the critical value $n_{cr}(m)$ if $m$ is too large, because that estimate then does not satisfy $n_{cr}(m)\gg m$. We will demonstrate this in sec. \ref{sec:eightpoint}.

\subsection{Truncating the leading asymptotic series}
\label{sec:asymptotics}

Notice that the leading $O(\partial^{2n})$ dependence \eqref{twoloop2m}  coincides with the expansion of a polylogarithm:
\be\label{polpow} -\text{Li}_\varsigma(-x) = -\sum_{n=1}^\infty \frac{(-x)^n}{n^\varsigma}\,,
\ee
in the regime where $x=c_m>0$. (In our case we also have $\varsigma=m+2-r-d/2$.) If, for the moment, we treat the leading terms as a formal series on their own, it is natural to
define the exact result for that series to be the left hand side: $-\text{Li}_\varsigma(-x)$. Indeed, for $m=1,2$, and $m=3$ provided $d<10-2r$ (\cf fig. \ref{fig:derivs}), the series converges; however we will also define the exact result for the series to be $-\text{Li}_\varsigma(-x)$ for all cases including those for $m\ge3$ where the series does not converge.

Now it takes just a short digression to prove that in general the most accurate estimate we can get from the series \eqref{polpow}, is gained by truncating it just short of the smallest term. This property follows because successive partial sums always bracket the exact result. To prove this, we use the integral expression:
\be 
-\text{Li}_\varsigma(-x) = \frac{1}{\Gamma(\varsigma)}\int^\infty_0\!\!\!\!dt\,\frac{\ x\,t^{\varsigma-1}}{\text{e}^t+x}\,,
\ee
(valid for $\varsigma>0$).
It is easily seen that the power series representation \eqref{polpow} follows from expanding:
\be 
\frac{x}{\text{e}^t+x} = \frac{x\,\text{e}^{-t}}{1+x\,\text{e}^{-t}} = -\sum^\infty_{n=1} \text{e}^{-nt}(-x)^n\,.
\ee
Resumming this geometric series starting from the $(N+1)^\text{th}$ term onwards, and substituting the result back into the integral, then establishes that
\be\label{exactR} -\text{Li}_\varsigma(-x) = -\sum_{n=1}^N \frac{(-x)^n}{n^\varsigma} - \frac{(-x)^{N+1}}{\Gamma(\varsigma)}\int^\infty_0\!\!\!\!dt\,\frac{t^{\varsigma-1}\text{e}^{-Nt}}{\text{e}^t+x}\,.
\ee
The second term on the right-hand side is thus an exact integral expression for the remainder after summing over the first $N$ terms. Substituting
\be \frac{1}{\text{e}^t+x} < \text{e}^{-t}\ee
(which holds since $x>0$) we see immediately that this remainder is of the same sign as the next term, only smaller. Therefore the exact answer for $-\text{Li}_\varsigma(-x)$ always lies between the sum to the $N^\text{th}$ term and the sum to the $(N+1)^\text{st}$, and thus, barring numerical flukes, the smallest error will be achieved by truncating the series just short of the smallest term.

This proves that the leading $O(\partial^{2n})$ terms \eqref{twoloop2m} considered as a series on their own, form an asymptotic series, in the sense described in the Introduction.  This is so, even though the series does not have the typical asymptotic form ``$n^\gamma n!\,  g^n$,'' that is  treated in the literature. Interestingly however, as we will see in sec. \ref{sec:gen}, it does take this form for other less well behaved cutoffs. 

Of course the leading terms here form an asymptotic series as defined in the Introduction, sec. \ref{sec:intro}, only 
if the parameters $\varsigma$ and $x$ are in the right regime. They are always in the right regime for infinitely many vertices, once $m$ is large enough, \cf sec. \ref{sec:eightpoint}. But we have already seen some exceptions. Firstly, for the two-point and four-point vertices we have $x<1$, so the series is not asymptotic but convergent, the most accurate result being the sum to $O(\partial^{\infty})$. Secondly, for the six-point vertex we have $x=1$, and thus the series is never asymptotic. Instead it either converges ($\varsigma=5-r-d/2>0$), oscillates without convergence ($\varsigma=0$), or diverges ($\varsigma<0$), as illustrated in fig. \ref{fig:derivs}. Finally, at fixed $m>3$, if $r$ and/or $d$ are large enough to make $\varsigma\le0$, these expansions are divergent and no longer asymptotic, because the smallest term in the series is now the first one ($n=1$) in the series.

\subsection{Watson's lemma}
\label{sec:watson}

The above proves that, apart from the exceptions just outlined, the leading $O(\partial^{2n})$ contributions \eqref{twoloop2m} to the two-loop diagrams of fig. \ref{fig:twoloop2m-2}, considered as a series on their own, form asymptotic expansions. But this is also true as a general non-perturbative statement for the derivative expansion of the functional renormalization group, in any dimension $d>0$. If the derivative expansion diverges, provided the analog of the $x$ parameter of sec. \ref{sec:asymptotics} is in the right regime, the resulting series is asymptotic, and as such results of some accuracy can still be extracted.  This follows from Watson's lemma \cite{Watson:lemma,Miller:aaa}. The lemma in particular implies that
\be\label{Watson} F(g)=\int^\infty_0\!\!\!\!\!\!du\ \text{e}^{-u/g}\, u^\sigma f(u)\ee
has an asymptotic expansion in $g>0$ if $f(u)$ is bounded and Taylor expandable about the origin $u=0$, and $\sigma>-1$. Since $K_\La(q)$  is just a Gaussian, \cf \eqref{Kexp}, all the quantum corrections we can get from the flow equation \eqref{ERG} can be cast in this form as follows. The vertices are Taylor expandable in $q^\mu$ and external momenta, thanks to the smooth infrared cutoff. Furthermore, appropriately defined,  the integrands that multiply $K_\La(q)$, derived from the flow equation \eqref{ERG}, are bounded as functions of $q$, in fact vanish as $q\to\infty$. Averaging over the angles for $q^\mu$ first, leaves only the integral over $|q|$, with external momenta combining into invariants amongst themselves. Then casting the remaining integral over $|q|$ as over $u=q^2/\La^2$ instead, maps it into the required form \eqref{Watson} with $g=1$ and $\sigma=\frac{d}{2}-1$. 

To verify this in the current example, note that the derivative expansion is formed from the loop-integral \eqref{qintp} over a one-loop vertex $\Gamma_1(q,-q,p,-p,0^{2m-2})$ where\footnote{As below \eqref{qpdidentity}, we have relabelled $m\mapsto m+1$.} we are interested in particular in the cases $2m\geq8$. These one-loop vertices are indeed Taylor expandable (in $q$, and $p$) and are bounded, in fact vanish, as $q\to\infty$. Averaging over the angles for $q^\mu$ first, converts the vertex effectively into a function of $q^2$ and $p^2$ separately, and then casting the remaining integral over $|q|$ as above, maps the integral into the required form \eqref{Watson}. 

As we have seen for the other vertices, there can also be exceptions in the sense that the series can be convergent (in which case summing to just before the smallest term means summing to infinity) or, since we are stuck with $g=1$ and thus have no ability to tune this parameter, in unfortunate cases it may lie outside the useful regime, so that in that case the smallest term in fact coincides with the first one in the series.

\subsection{Asymptotic behaviour of two-loop eight-point vertices and higher}
\label{sec:eightpoint}

Now we analyse in more detail the leading order  of the derivative expansion for two-loop  $(2m\geq 8)$-point vertices. In particular we test to what extent we can get accurate answers from the derivative expansion, even though it diverges for these vertices. For $(2m\geq 8)$-point vertices the exception (\ref{subst},\ref{substex}) only appears in high dimensions: $d\ge6$. Irrespective of whether the exception exists or not, writing the magnitude of the leading order $O(\partial^{2n})$ contribution to these vertices as $\exp -A(n)$, we see that
\be \frac{\partial A}{\partial n} = \frac{2m+4-(2r+d)}{2n}-\ln c_m \,.\ee
Recall from sec. \ref{sec:twoloops} that in order to extract an accurate answer from the six-point vertex, the leading order derivative expansion tells us that we require $2r+d\le9$. Since we also have $2m+4\ge 12$ for the $(2m\geq 8)$-point vertices,  we confirm that for small enough $n$, $\partial A/\partial n$ is positive and thus the leading asymptotic form of the $O(\partial^{2n})$ contributions initially decreases with $n$. On the other hand, we already know that the contributions eventually increase with increasing $n$ (and can likewise readily confirm this from the above equation).
The minimum size contribution is then reached, according to this leading asymptotic formula, at $\partial A/\partial n = 0$, \ie such that
\be\label{ncrm} n = n_{cr}(m) = \frac{2m+4-(2r+d)}{2\ln c_m}\,.\ee
If the correction described below \eqref{qintp} is the dominant contribution to $\Gamma_2(p,-p,0^{2m-2})|_{\partial^{2n}}$ and if this value of $n_{cr}(m)$ is large enough to trust the leading asymptotic formula, which in particular requires $n_{cr}(m)\gg m$, then we will  have found the value of $n$ that gives the smallest term in the full two-loop computation, and truncating the derivative expansion at $n=n_{cr}(m)-1$ will then give the most accurate estimate to two loops for this $2m$-point $2r$-derivative operator. 

\begin{figure}[ht]
\centering
\includegraphics[scale=0.155]{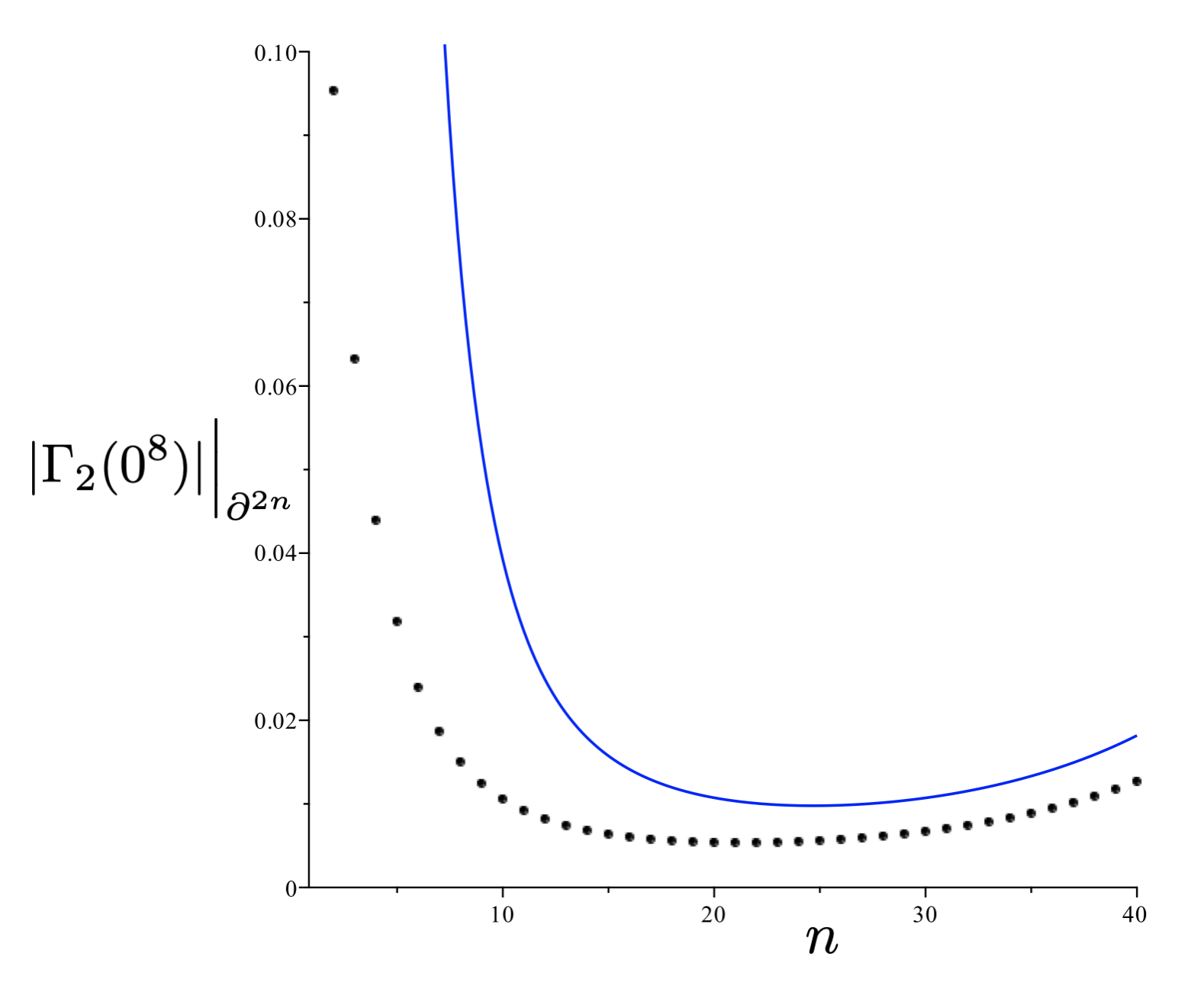}
\includegraphics[scale=0.155]{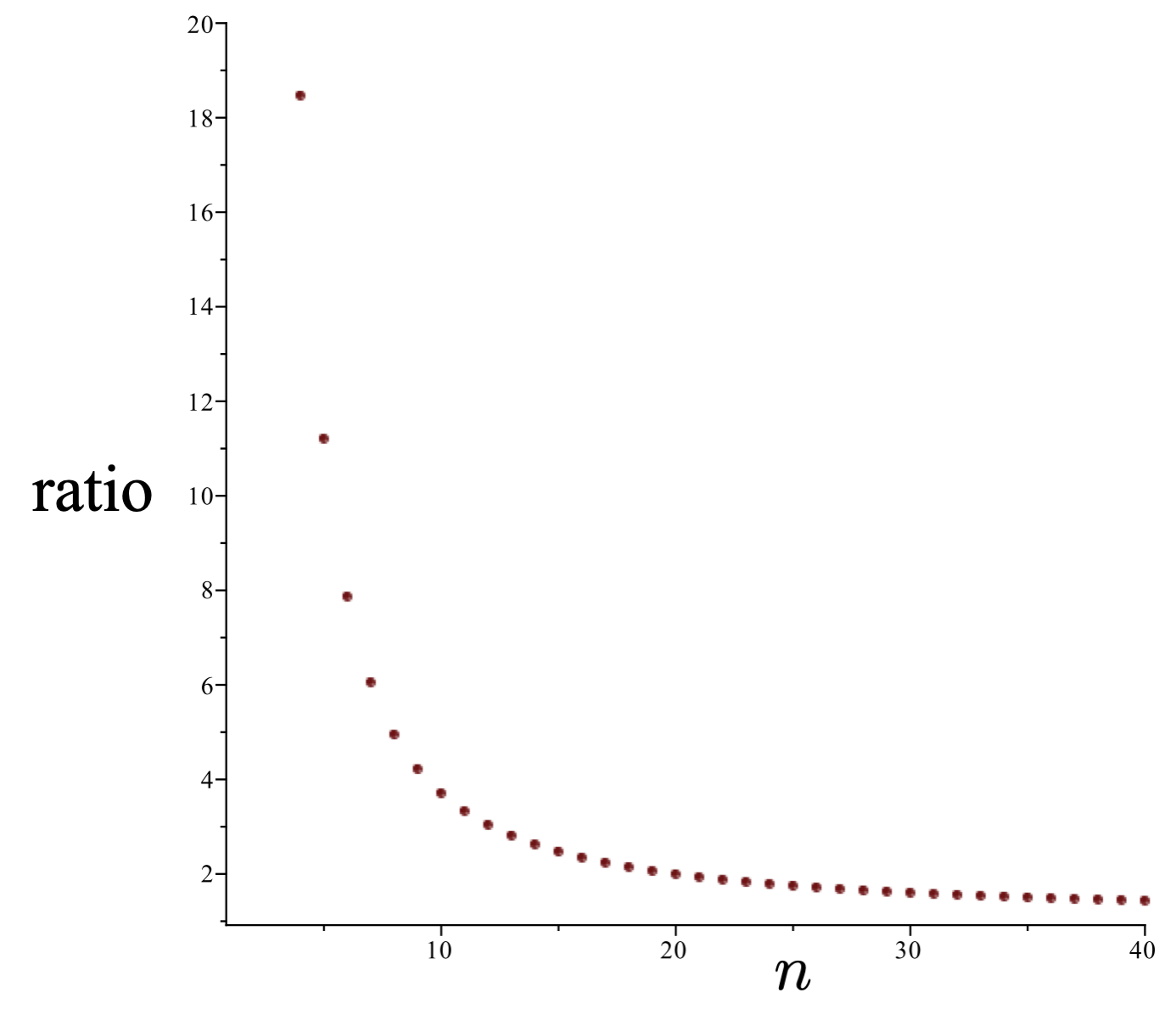}
\caption{The graph on the left displays the magnitude of derivative expansion contributions, via  \eqref{qint0}, to the eight-point vertex in the two-loop effective potential for three dimensional $\lambda\ph^4$ theory, from $O(\partial^2)$ up to $O(\partial^{80})$. The numerical value is the coefficient of $\lambda^5/\La^6$. It is plotted in blue using the asymptotic formula \eqref{2loop2mi} and compared to the exact formula \eqref{2l2mexi} displayed as black dots. The graph on the right plots the ratio of the asymptotic formula to the exact formula.
}
\label{fig:plot8}
\end{figure}

By inspection, we see that $m/n_{cr}(m)$ is smallest when $m=4$ so that we get the smallest value for $\ln c_m$, namely $\ln c_m=\ln\frac65 = 0.1823$. To test how good the asymptotic formula actually is, we compare the exact and asymptotic formulae, \eqref{2l2mexi} and \eqref{2loop2mi} respectively, for the important example of $d=3$ and $r=0$ (as relevant to the equation of state of the three dimensional Ising model fixed point). Since we set $r=0$, these formulae are for \eqref{qint0} (or rather its integrated version) and thus expected to be the dominant contribution. The results\footnote{The integrals $I^n_{5,1}$ and $I^n_{5,2}$ were found in closed form after a long computation using Maple.} are plotted in fig. \ref{fig:plot8}. 

Note that only the magnitude is plotted. We remind the reader that, since the sign of the derivative expansion contribution oscillates as $(-1)^n$, the summed result will display (apparent) convergence so long as the magnitude drops with increasing $n$.

We see that the asymptotic formula is increasingly accurate for larger $n$, as expected. The leading asymptotic result \eqref{ncrm} predicts that the minimum contribution occurs at $n_{cr}(4)=24.68$ whilst the exact result (which is very flat around the minimum) actually has its minimum at $n_{cr}(4)=21$. Thus the leading asymptotic result predicts that the derivative expansion continues to converge up to $O(\partial^{48})$ whilst the exact result (for this set \eqref{qint0} of contributions) gives convergent results up to $O(\partial^{40})$.

The discrepancy between these two estimates is na\"\i vely of the order we might expect, since we expect the leading asymptotic formula to have multiplicative corrections $\sim m/n$, and in this case $m/n\sim 20\%$. However the asymptotic estimate for the magnitude itself, breaks down more quickly at lower $n$ than one might have expected. In particular a multiplicative correction of almost 1/2 is needed to reproduce the exact answer at $n=21$, whilst at $n=4$ where one might have hoped for errors of $O(1)$, the asymptotic answer is actually 18 times larger than the exact answer.

Of course this numerical analysis is only for the contribution \eqref{qint0} to the eight-point vertex ($m=4$). To obtain estimates more generally of the limits of accuracy of the derivative expansion \eg for the full equation of state, we would need an estimate for $n_{cr}(m)$ that is valid for all $m\ge4$. 

As $m$ increases, the asymptotic formula \eqref{ncrm} predicts that $n_{cr}$ falls and reaches a minimum at $\partial n_{cr}(m)/\partial m =0$ after which it increases with increasing $m$. Since the value $m:=m_{cr}$ that gives this minimum $n_{cr}$, turns out to be large, we can neglect the $1/(m+1)^2$ correction in \eqref{cm}, and use $c_m=(m+1)/4$ for both even and odd $m$. Then, solving for $\partial n_{cr}(m)/\partial m =0$ gives:
\be\label{mneglect} \ln\frac{m+1}{4}= 1-\frac{d+2r-2}{2(m+1)}\,.\ee
Again, for $m$ large we can neglect the second term on the right, and we thus find that the vertex with the smallest $n_{cr}(m)$ is such that 
\be\label{mapp} m=m_{cr} \approx 4\,\text{e}-1 = 9.9 \approx 10\,, \ee
\ie  it is approximately the twenty-point vertex that can be expected to have the smallest length trustable derivative expansion, independent of dimension $d$ and number of derivatives $r$. And indeed for interesting values of $d$ and $r$, the neglected correction in \eqref{mneglect} is approximately a $10\%$ effect, whilst the neglected difference between even and odd expressions for $c_m$ is an error of $1 \%$. Note that at this level of approximation \eqref{mneglect} implies that $\ln c_{m_{cr}} =1$. Then from \eqref{ncrm} we find the minimum critical value of $n$ is:
\be\label{ncra} n_{cr}(m_{cr})  \approx n_{cr}(10) \approx 12 - r-\frac{d}{2}\,.\ee
According to this formula the derivative expansion can be trusted to very high order for all two-loop vertices. For example for the effective potential (and thus the equation of state) in $d=3$ and $d=4$ dimensions the derivative expansion continues to converge up to $O(\partial^{18})$ and in $d=2$ dimensions up to $O(\partial^{20})$.

Unfortunately, upon rearranging \eqref{ncrm}, it is clear that as $m$ increases, $m/n_{cr}(m)$ also increases. Therefore the leading asymptotic result around $n=n_{cr}(m)$ becomes increasingly unreliable with increasing $m$. In particular at $m=m_{cr}\approx 10$ we have that $m_{cr}/n_{cr}\sim 1$ and thus, especially given the already poor behaviour we saw above, we can no longer trust these estimates. Furthermore, of course, this estimate is only for the two-loop contribution and thus, being only part of the answer, cannot be a reliable estimator for the full derivative expansion.

\subsection{Subleading two-loop contributions}
\label{sec:othertwo}

\begin{figure}[ht]
\centering
\includegraphics[scale=0.15]{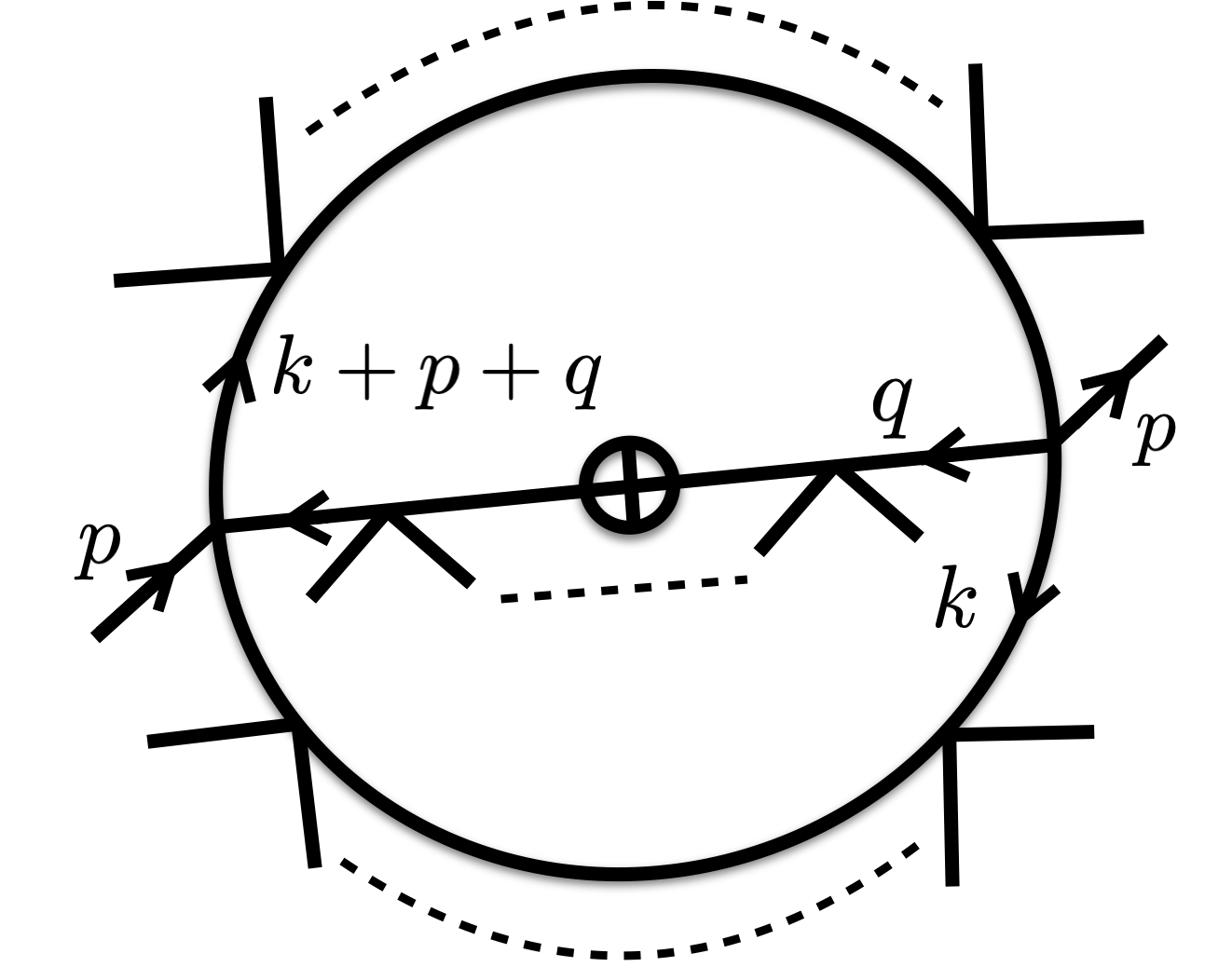}
\caption{A two-loop Feynman diagram contribution to the flow of the $(2m)$-point vertex $\Gamma_2(p,-p,0^{2m-2})$, of the form of fig. \ref{fig:twoloop2m-2}, but where $2s$ of the external legs are introduced through vertices `decorating' the new  $q$-propagator line.}
\label{fig:twoloop2ms}
\end{figure}

We have argued that at two loops, contributions of the form of fig. \ref{fig:twoloop2m-2} dominate at large $O(\partial^{2n})$, \ie give the leading convergence properties at two loops as $n\to\infty$. This was based on identifying the contribution to the two-loop $\beta$ function with the weakest convergence, recognising that this property follows from using the one-loop operator with the greatest mass dimension, and then generalising this observation to propose the contribution fig. \ref{fig:twoloop2m-2} as the least convergent for two-loop $2m$-point vertices. Now we further support this claim. Suppose instead we compute the contribution of fig. \ref{fig:twoloop2ms}, where $2s$ of the external legs ($s\le m-1$) now emanate from the new propagator line introduced at two loops. This corresponds to starting with $2s$ fewer legs in the one-loop vertex which, according to the arguments above, should result in improved convergence for the derivative expansion.

To get this contribution, all we need to do is insert a factor of \eqrf{p75-76}
\be \Delta_{IR}^s(q) = \frac1{\La^{2s}}  \int^1_0\!\!\!da_1\cdots da_s\, \text{e}^{-A\,q^2/\La^2}\,,\qquad A= \sum_{i=1}^sa_i\,, \ee
into the quantum correction \eqref{qintp}, and shift $m\mapsto m-s$.  The net effect of this insertion is that the $q$-integral \eqref{qpdidentity} is now replaced by:
\besp 
\int^1_0\!\!da_1\cdots da_s\int\!\!\frac{d^dq}{(2\pi)^d}\, (q+p)^{2n}\, \text{e}^{-(1+A)\,q^2/\La^2} =\\ \frac{\La^{d+2n}}{(4\pi)^{d/2}}\int^1_0\frac{da_1\cdots da_s}{(1+A)^{n+d/2}}\sum_{r=0}^n \frac{\Gamma(n+d/2)}{\Gamma(r+d/2)}\binom{n}{r}\left(\frac{p}{\La}\right)^{2r}(1+A)^r\,.
 \eesp
At large $n$, these Schwinger integrals are dominated by the region $a_i\to0$. To leading order we need only this boundary contribution, which makes the integrals straightforward, and thus we see that the sole effect is to introduce a factor of $1/n^s$ into the previous result \eqref{twoloop2m}. Remembering now the compensating shift $m\mapsto m-s$, we see that for the contribution of fig. \ref{fig:twoloop2ms} we have:
\be
\label{twoloop2ms}
\partial_\La\Gamma_2(p,-p,0^{2m-2})|_{\partial^{2n}} \ni C_{m-s}\La^{2d-2m-5}\sum_{r=0} \frac{n^{r-m+d/2-2}(-c_{m-s})^n }{r!\,\Gamma\!\left(r+\frac{d}{2}\right)}\,\left(\frac{p^2}{\La^2}\right)^r
+\cdots\,.
\ee
Apart from the overall prefactor (which does not alter the $n$ dependence) the only change is to replace $c_m$ by $c_{m-s}$ which means that the exponential growth in $n$ has been reduced (as is clear from their definition \eqref{cm}). Thus we see indeed that this contribution has improved convergence compared to the contributions in fig. \ref{fig:twoloop2m-2}. 
In particular, no matter how large $m$ is, if $m-s\le2$, then the derivative expansion actually converges for these contributions.

\subsection{Three loops}
\label{sec:threeloop}

\begin{figure}[ht]
\centering
\includegraphics[scale=0.15]{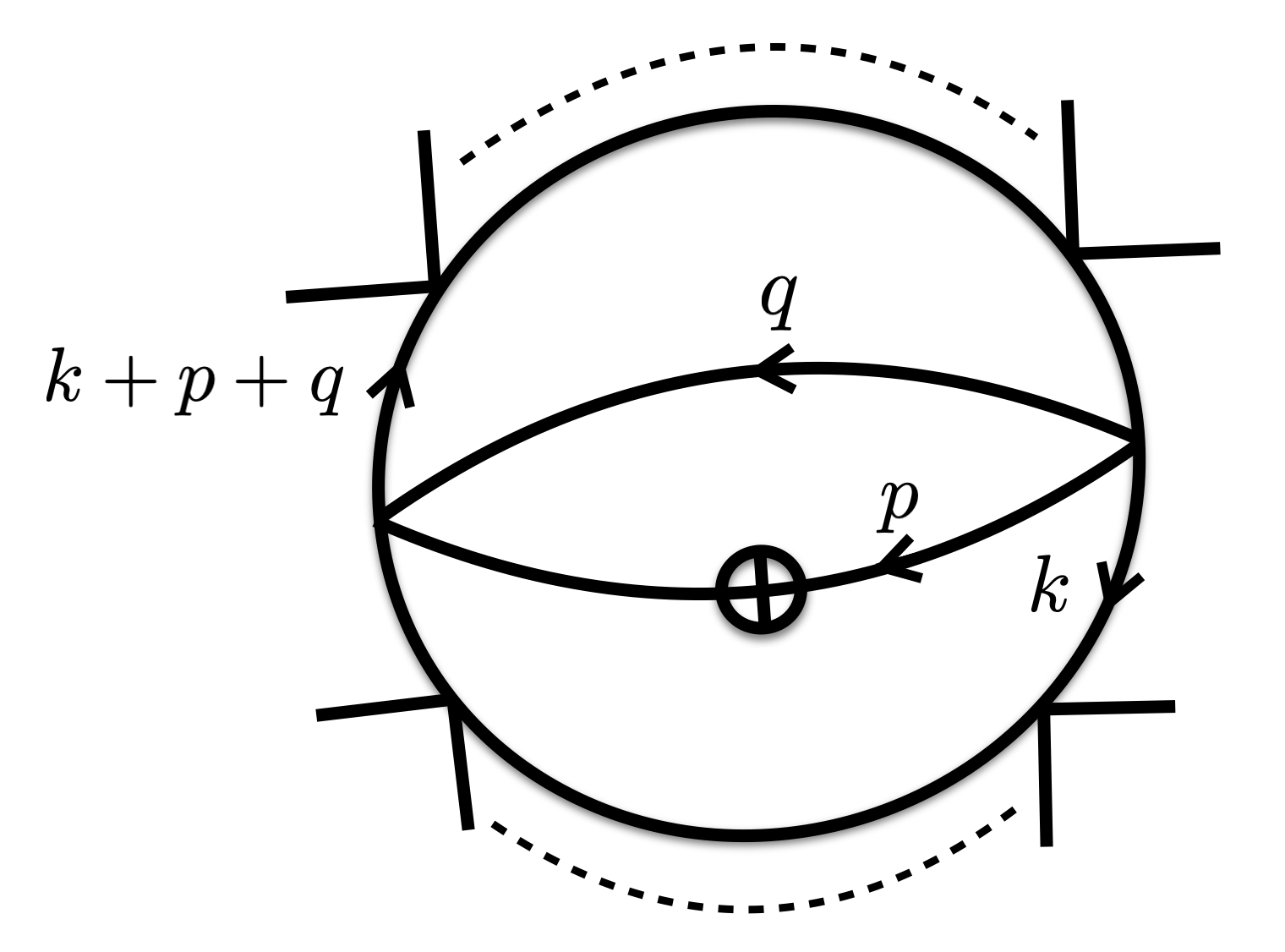}
\caption{A three-loop Feynman diagram contribution to the flow of the $(2m-2)$-point vertex $\Gamma_3(0^{2m-2})$, formed by tying together the $p$-dependent external legs  in the contribution to the two-loop $2m$-point vertex illustrated in fig. \ref{fig:twoloop2m-2}, and integrating them over $K_\La(p)$ (indicated here by the crossed circle).  The unlabelled external legs carry zero momentum.}
\label{fig:threeloop}
\end{figure}

We have seen that the derivative expansion at one loop (which coincides with a Taylor expansion in external momenta) has an infinite radius of convergence, coefficients of the $O(\partial^{2n})$ terms dropping factorially with $n$. This convergence is weakened or replaced by asymptotic series behaviour at two loops.  Around eqn. \eqref{momint} and then in more detail in the second paragraph below eqn. \eqref{twlotwint}, we  identified the reason for that. It is a generic feature  that on substituting a one-loop vertex expanded in powers of momentum, into a loop momentum integral, the cutoff loop integral over powers of loop momentum supplies a numerical factor (\ref{momint}) that in large part nullifies the decay in the Taylor expansion coefficients. We saw that at work in the previous sections, where we took the derivative expansion \eqref{largen2m} of the one-loop $2m$-point vertex and subjected it to an integral over the cutoff as in fig. \ref{fig:twoloop2m-2}, resulting in the asymptotic $O(\partial^{2n})$ behaviour \eqref{2loop2mi}. At three loops, we can subject the \emph{same} one-loop Taylor expansion to one further integral over the cutoff via: \eqrf{41.6}
\be 
\label{dGthreez}
\partial_\La \Gamma_3(0^{2m-2})|_{\partial^{2n}} \ni -\frac12 \int \frac{d^dp}{(2\pi)^d} K_\La(p)\, \Gamma_2(p,-p,0^{2m-2})|_{\partial^{2n}}\,,
\ee
since one such contribution involves integrating over the $p$ dependence in (\ref{2l2mexi},\ref{substex}) forming the Feynman diagram in fig. \ref{fig:threeloop}.  

We need expressions (\ref{2l2mexi},\ref{substex}) rather than their asymptotic versions (\ref{2loop2mi},\ref{subst}) so that they apply when $r$ is not necessarily much less than $n$. Notice that this does \emph{not} mean that we are now effectively taking a Taylor expansion in $p^{2r}$ with finite radius of convergence, and multiplying it by $\sim r!$ as in (\ref{momint}). Although at two loops the $O(\partial^{2n})$ expansion for the coefficients has weakened convergence or even diverges, as a function of $r$ they still drop factorially as is clear from \eqref{2l2mexi}, and also \eg the resummed expression \eqref{2l2} for the self-energy.

As at two loops, we expect the contribution \eqref{dGthreez} to be the weakest convergence obtainable, in this case also because we have now run out of legs on the same $\ph^4$ vertex to sew together in a way that subjects the same one-loop Taylor expansion to further loop momentum integrals. Other derivative expansion contributions of form \eqref{dGthreez} would have to come from loop integrals over different one-loop Taylor series in momenta, and we expect such contributions to have improved convergence compared to the above case as we explain below. Finally, recall that we have already seen in sec. \ref{sec:othertwo} that introducing $\ph^4$ vertices beyond those already at one loop also significantly improves the convergence. Thus we believe that the contribution in fig. \ref{fig:threeloop} gives the weakest convergence obtainable at three loops and even when compared to the leading contributions at higher loop order.

For the integral of (\ref{2l2mexi},\ref{substex}) over
$K_\La(p)$, we just need the generalisation of the four dimensional identity \eqref{momint}:
\be 
\label{momind}
\frac12\int\!\!\frac{d^dp}{(2\pi)^d}\, K_\La(p)\, p^{2n} = \frac{\La^{d-3+2n}}{(4\pi)^{d/2}} \frac{\Gamma(n+d/2)}{\Gamma(d/2)}\,,
 \ee
which follows immediately from the definition \eqref{Kexp} and the $r=0$ term in the identity \eqref{qpdidentity}. Using $n\gg d,m$, we get:
\besp 
\label{G30}
\partial_\La \Gamma_3(0^{2m-2})|_{\partial^{2n}} \ni \frac12 \frac{C_m\La^{3d-2m-7}(-c_m)^n}{\Gamma\!\left(\frac{d}{2}\right)(4\pi)^{d/2}n^{m+1}}\Bigg\{ \!\sum^n_{r=0\atop r\ne d-m-2}\frac{\Gamma\!\left(n+\frac{d}{2}\right)}{r!(n-r)!(m+r+2-d)}\\ 
+2\,\theta(d-m-2)
\frac{\Gamma\!\left(n+\frac{d}{2}\right)\Gamma\!\left(\frac{3d}{2}-m-2\right)}{(d-m-2)!(n-d+m+2)!}\ln\frac{\La}{\mu}
\Bigg\}\,,
\eesp
where the Heaviside theta function is defined to satisfy $\theta(0)=1$. This latter term follows from the exceptional case (if it exists) where $r=d-m-2$. The large $n$ dependence of the sum may be evaluated by noticing that it is captured by the $a\to1$ part of the integral:
\beal 
\sum^n_{r=0\atop r\ne d-m-2}\frac{\Gamma\!\left(n+\frac{d}{2}\right)}{r!(n-r)!(m+r+2-d)} &= n^{d/2-1}
\int^1\!\!\!\! da\, (1+a)^n a^{m+1-d}\quad+\cdots\,,\nn\\ 
&= 2^{n+1} n^{d/2-2}+\cdots\,,
\eeal
where in the last step as usual we have set $a=1-\alpha$ and kept first order in $\alpha$ only in the $n^\text{th}$ power. In comparison the exceptional term in \eqref{G30} $\sim n^{3d/2-m-3}$, is always subleading. Thus altogether we have shown that:
\be 
\label{G30as}
\partial_\La \Gamma_3(0^{2m})|_{\partial^{2n}} \ni \frac{C_{m+1}\La^{3d-2m-9}}{\Gamma\!\left(\frac{d}{2}\right)(4\pi)^{d/2}}n^{d/2-m-4}(-2c_{m+1})^n+\cdots
\ee
(where we have shifted $m\mapsto m+1$ so that it again counts half the number of legs in the computed vertex).

The above contributes at three loops to the effective potential.
Notice that in comparison to the two-loop contribution to $\partial_\La\Gamma_2(0^{2m})|_{\partial^{2n}}$, \viz the $r=0$ part of \eqref{twoloop2m}, the power of $n$ has improved by a factor $1/n^2$ but the exponential in $n$ dependence has worsened: $-c_m\mapsto -2c_{m+1}$. As a result the derivative expansion now diverges for every $m\ge1$, but its leading order terms again form an asymptotic series. This is so because we require $d\le9$ in order for the two-loop six-point vertex to have a useful derivative expansion as discussed in sec. \ref{sec:twoloops} and thus the power of $n$ in \eqref{G30as} is negative for all $m\ge1$. This in turn implies that the contribution above initially falls in magnitude with increasing $n$ before reaching a minimum and then growing exponentially with $n$. 

Note that for the four-point vertex, we summed the two-loop contributions (\ref{2l2mexi},\ref{substex})  for the six-point vertex over $r=0$ to $r=n$, all with the same sign, replacing the power $p^{2r}$ with the factorially growing $\Gamma(r+d/2)$ via \eqref{momind}. Nevertheless we have ended up with an asymptotic series, potentially providing the conditions for accurate estimates to be extracted from the derivative expansion for this four-point vertex at three loops. This may look counter-intuitive, given that for $r>4$ the six-point vertex has a divergent derivative expansion with no asymptotic regime,  \cf \eqref{2loop2mi} and fig. \ref{fig:derivs}. But there is no contradiction. The series for the six-point vertex \eqref{2loop2mi} with $r>4$, diverges as a power of $n$, but the asymptotic series \eqref{G30as} eventually diverges exponentially with $n$. Despite this, the latter series lies in a potentially useful regime that displays asymptotic behaviour, even though the former does not. 

Less dramatically perhaps but still worth remarking, higher point vertices have two-loop contributions which are asymptotic expansions, and summing over these, all with the same sign, replacing the power $p^{2r}$ with the factorially growing $\Gamma(r+d/2)$, results at three loops in another  potentially useful asymptotic expansion, but one whose eventual exponential growth is faster than at two loops.

Going through the same steps as in sec. \ref{sec:eightpoint}, we can compute that the smallest term magnitude is reached at
\be\label{ncr3} n=n_{cr}(m) = \frac{m+4-d/2}{\ln(2c_{m+1})}\,.\ee
For example we find for $d=3$, and $m=1$ that the minimum value is reached at $O(\partial^{24})$ ($n_{cr}=12.1$) where an accuracy of $1/n_{cr}\approx 8\%$ might be expected. 

For larger $m$, the same simplified approximations as before, again turn out to work, accurately reproducing the results from \eqref{ncr3}. Thus we find that the minimum $n_{cr}(m)$ occurs when $\ln(2c_{m+1})\approx 1$ \ie at $m=m_{cr}\approx 3.4$ and this corresponds to $n_{cr}=n_{cr}(m_{cr})\approx 7.4 - d/2$. This suggests that for the effective potential, the derivative expansion will continue to furnish increasingly accurate results up to approximately $O(\partial^{10})$ for dimensions $d=3,4$ and $O(\partial^{12})$ for  $d=2$.

As before however,  \eqref{ncr3} tells us that the accuracy $\sim m/n_{cr}(m)$ decreases as $m$ increases, so we would have to go beyond leading order in $n$ to find an estimate for the three-loop $n_{cr}(m)$ that is reliable for all $m$, and in particular at $m\approx 3.4$ where the na\"\i ve estimate already suggests errors of $\sim 60\%$.

\subsection{Subleading three-loop contributions}
\label{sec:otherthree}

\begin{figure}[ht]
\centering
\includegraphics[scale=0.15]{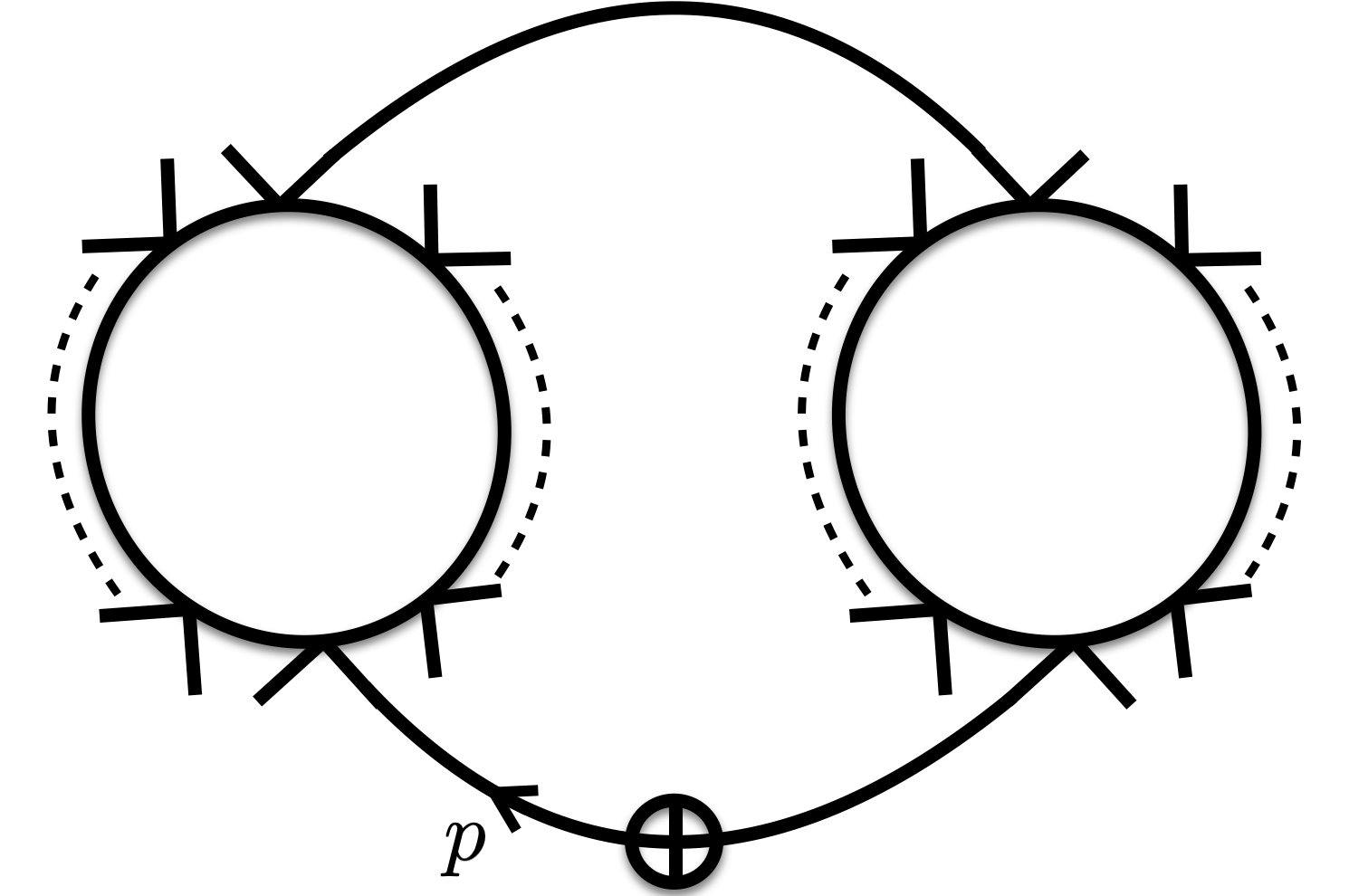}
\caption{A three-loop Feynman diagram contribution where the loop momentum flows are not concentrated through the same $\ph^4$ vertex.}
\label{fig:threeloopalt}
\end{figure}

Now let us return to the point made in the previous subsection, that other derivative expansion contributions arise from loop integrals over different one-loop Taylor series in momenta, leading to significantly improved convergence compared to the above case. Consider for example the three-loop contribution
\besp \partial_\La \Gamma_3(0^{2(m_1+m_2)-4})|_{\partial^{2n}} \ni \\
C_{m_1,m_2}\sum_{n_1=0}^n \int \frac{d^dp}{(2\pi)^d} K_\La(p)\, \Gamma_1(p,-p,0^{2m_1-2})|_{\partial^{2n_1}}\Delta_{IR}(p)\,\Gamma_1(p,-p,0^{2m_2-2})|_{\partial^{2(n-n_1)}}
\,,\eesp
where $C_{m_1,m_2}$ is some $n$-independent combinatorial factor. This is illustrated in fig. \ref{fig:threeloopalt}. Once integrated up with respect to $\La$ we would arrive at the same result by taking the quantum correction of  fig. \ref{fig:twoloop2ms} and, at three-loop order, attaching a further `decorated' propagator to the $q$-dependent propagator line.

The $p$ integral can be cast in the form
\be \int \frac{d^dp}{(2\pi)^d}\, \text{e}^{-p^2/\La^2} \left(1- \text{e}^{-p^2/\La^2}\right) (p^2)^{n-1}
\propto \Gamma(n-1+d/2) \left(1-2^{1-n-d/2}\right) = \Gamma(n-1+d/2)+\cdots\,,
\ee
where we display only the $n$-dependent contributions and recognise that the power of $2$ is subleading. The leading contribution at large $n$ from the $\Gamma_1(p,-p,0^{2m_i-2})$ contributions, will come from maximising both of them \ie by setting $m_1=m_2=m/2+2$ and $n_1=n_2=n/2$.\footnote{This assumes even $m$ and $n$ but nearest integer choices for odd values give the same leading asymptotics.} Then substituting \eqref{largen2m} with these values and using Stirling's formula $n!\sim \sqrt{2\pi n}\, (n/\text{e})^n+\cdots$ on the above and the factorials, we find finally that the leading $n$ dependence of this contribution to $\partial_\La \Gamma_3(0^{2m})|_{\partial^{2n}}$ is
\be 
n^{d/2-m-9/2} (-2 c_{m/2})^n\,.
\ee
This asymptotic behaviour is indeed a significant improvement on \eqref{G30as}. Thus we see that, as advertised, contributions that combine different derivative expansions in a loop integral actually result in better behaved asymptotic expansions compared to integrating again over the same expansion as in fig. \ref{fig:threeloop}. The power of $n$ is improved by a factor $1/\sqrt{n}$, but the big improvement is in the exponential dependence which, because the derivative expansion was split across two contributions, results in the replacement $c_m\mapsto c_{m/2}$. 

Indeed for $m/2$ odd, we have by \eqref{cm} that $2c_{m/2}=c_{m+1}$ and thus this contribution is comparable to the two-loop example, the $r=0$ part of \eqref{twoloop2m}: slightly worse exponentially but the power of $n$ is improved by $n^{-5/2}$. Going through the same exercises as before, one finds a marginal \emph{improvement} in its asymptotic behaviour compared to the two-loop case, with convergence at least up to $O(\partial^{22})$. However, as before we expect subleading contributions to be significant at larger $m$.

\section{General cutoffs}
\label{sec:gen}

So far all our investigations have been with the exponential cutoff \eqref{Rspecial}. We claimed at the end of sec. \ref{sec:expc} however that this gave the best behaviour for the derivative expansion. We will now justify this claim. For concrete examples we use those cutoffs advocated in ref. \cite{Balog:2019rrg} that are differentiable to all orders (so that an all-orders derivative expansion exists \cite{Morris:1999ba})
namely the Wetterich regulator \cite{Wetterich:1992} $R_\La = W_\La$, and an exponential regulator $R_\La= E_\La$, where:
\beal W_\La(q) &= \frac{\alpha\, q^2}{\text{e}^{\,q^2/\La^2}-1}\,,\\
         E_\La(q) &= \alpha \La^2 \, \text{e}^{-q^2/\La^2}\,.
\eeal 
In ref. \cite{Balog:2019rrg,Wetterich:1992} a wavefunction renormalization factor $Z^0_\La$ is also included but we do not need that here. The parameter $\alpha$ has been used to tune the results, order by order in the derivative expansion, according to a ``principle of minimal sensitivity''. The resulting values of $\alpha$ start around $5$ ($E_\La$) or $6$ ($W_\La$) at $O(\partial^0)$ but fall rapidly towards $\alpha \gtrsim 1$ by $O(\partial^6)$ \cite{DePolsi:2022wyb}. Note that should the trend continue and $\alpha\to1$ at high derivative expansion order, then $W_\La$ coincides in this limit with the exponential cutoff \eqref{Rspecial} we have been using throughout the paper until now.

When $\alpha\ne1$ in the $W_\La$ case, and in general for $E_\La$, the resulting dependence on $q$ is \emph{not} analytic everywhere in the complex plane except at infinity, and this leads, at one loop to momentum expansions that have a finite radius of convergence $r(\alpha)$,
\ie
\be \Gamma_1(p) = \sum_{n=0}^\infty a_n\, p^{2n}/\La^{2n}\,,\ee
where 
\be a_n= n^\gamma (-1)^n/r^n(\alpha)+\cdots\,,\ee 
instead of the infinite radius of convergence factorially decaying $a_n$ we had previously. Here we write $\Gamma_1(p)$ to stand generically for any one-loop vertex, scaled to be dimensionless, with dependence on a single external momentum $p$. For later use, we include the factor $n^\gamma$ that will generically be present in the large $n$ limit, where $\gamma$ is some number depending on the choice of cutoff \etc The ellipses stand for terms that are subleading in a large $n$ expansion, as before. The radius $r(\alpha)$ is set, roughly speaking, by the distance to the nearest singularity from $q^2\approx \La^2$ (roughly the average value of the loop momentum in a loop integral regulated by the cutoff).  For $W_\La$, we have that 
\be \Delta_{IR}(q,\La) = \frac1{q^2+W_\La(q)} = \frac1{q^2} \frac{\text{e}^{\,q^2/\La^2} -1}{\text{e}^{\,q^2/\La^2}+\alpha -1}\,,\ee
and for $\alpha>1$ as above, the closest pole is at $q^2 = \La^2\left\{\ln(\alpha-1)+i\pi\right\}$, which gives for example $r(6)\approx 3$. For $\Delta_{IR}(q,\La) = 1/(q^2+E_\La)$, the closest pole is at $q^2=\La^2 \omega(-\alpha)$, where $\omega$ is the Lambert W function on its principal branch. At $\alpha=1$, this lies at $q^2=\La^2 (-0.3181+1.337 i)$, and thus gives $r(1)\approx 2$.

We cannot compare these radii of convergence to the calculations done in practice  however, on the one hand because those were improved by changing $\alpha$ between orders \cite{Canet:2002gs,DePolsi:2022wyb}, and on the other hand the situation is anyway changed completely by the integration over these powers of momentum that takes place in forming the derivative expansion. At large order $O(\partial^{2n})$, this loop integration over powers of momentum $p^{2n}$, for example at two loops, explores only the high momentum tail of these cutoffs where the decay will be exponential, similar to before. Thus this integral multiplies the coefficients $a_n$ by a factorially growing factor as we saw before.
To see this more concretely note that, at large $n$, the integral 
\beal \frac12\int\!\!\frac{d^dp}{(2\pi)^d}\, K_\La(p)\, p^{2n} &= \frac1{(4\pi)^{d/2}\Gamma(d/2)}\int_0^\infty \!\!\!\!\!dp\, K_\La(p)\, p^{2n+d-1}\nn\\
&= \alpha\frac{\La^{d-3+2n}}{(4\pi)^{d/2}} \frac{\Gamma(n+d/2)}{\Gamma(d/2)}+\cdots\qquad\quad\  (R_\La=W_\La)\,,\nn\\
&= \alpha\frac{\La^{d-3+2n}}{(4\pi)^{d/2}} \frac{\Gamma(n-1+d/2)}{\Gamma(d/2)}+\cdots\qquad (R_\La=E_\La)\,,
\label{momgen}
 \eeal
is dominated by a peak at large $p$ where  the exponential suppression in $K_\La(p)$ finally wins out. The leading behaviour can thus be computed by steepest descents. However, knowing that it is dominated at large $p$, it is simpler to note that in that limit 
\beal K_\La(p) &\to  \frac{2\alpha}{\La^3}\,\text{e}^{-p^2/\La^2}\qquad\ (R_\La=W_\La)\,,\nn\\
&\to \frac{2\alpha}{p^2\La}\,\text{e}^{-p^2/\La^2}\qquad (R_\La=E_\La)\,,
\label{Klimits}
\eeal
which, by the corresponding expression for the exponential cutoff \eqref{Kexp} and its integral \eqref{momind}, yields the expressions above.

Thus at two loops we end up with high order terms in the derivative expansion behaving as $n^{d/2+\gamma'}n!(-1/r)^n+\cdots$, where $\gamma'=\gamma-1$, $\gamma-2$ for $W_\La$, $E_\La$ respectively. This behaviour is again that of an asymptotic series, in fact in the form that it is usually found in the literature, with $1/r(\alpha)$ playing the r\^ole of the small parameter. And indeed thanks to the limits \eqref{Klimits}, the conditions for Watson's lemma \eqref{Watson} still apply for the leading behaviour at large $n$. For $R_\La=E_\La$, they apply only for dimensions $d>2$ however, because in that equation we need now $-1<\sigma=\frac{d}{2}-2$. On the other hand for $W_\La$, the conditions apply for any dimension $d>0$ as before and furthermore, tuning $\alpha\to1$ sends $r(\alpha)\to\infty$ and thus in this case we even have small parameter control over the asymptotic series (although this is not $g$ in Watson's lemma \eqref{Watson}, which remains at $g=1$).

However, the coefficients now ultimately grow factorially, and thus much faster than the exponentially fast growth we had previously. Also, previously at two loops the two-point and four-point vertices had convergent derivative expansions (in any dimension) whereas now they are only asymptotic, although of useful form,  \ie with terms that decrease, as $n$ increases, down to some minimum at $n=n_{cr}$, and then grow (factorially fast) thereafter. For fixed $\alpha$, for $2m$-point vertices we can expect in general, as before, \cf \eqref{twoloop2m}, that we pick up a factor $\sim c_m^n$ but that also the power $\gamma'$ drops like $-m$. Thus as $m$ increases, even though eventually we will have $c_m/r>1$, the asymptotic series could continue to be of a useful form, thanks to the decreasing power of $n$. To decide whether this is really the case, and to decide to what extent the leading terms on their own are sufficient near $n_{cr}$ (\cf sec. \ref{sec:eightpoint}), would require a detailed study with these cutoffs. Given their more complicated analytic structure, that study would probably have to  proceed largely numerically.

\section{Discussion and Conclusions}
\label{sec:conclusions}

As we remarked in the beginning, if the derivative expansion were a convergent expansion then it would also have to converge in the loop expansion, since this is nothing but a Taylor expansion of the full non-perturbative result in Planck's constant, $\hbar$. Another way of seeing this is as follows. The right-hand side of the flow equation \eqref{ERG} takes the form of the one-loop correction \eqref{ERGone}, except that the classical action $\Gamma_0$ is replaced by the full non-perturbative Legendre effective action $\Gamma$ \cite{Wetterich:1992}. Thus we expect the derivative expansion of the left hand side, \viz $\partial_\La\Gamma$, to converge at best like that of the one loop correction. Indeed if we solve for the flow of $\Gamma$ starting at $\Gamma=\Gamma_0$ in the UV (ultraviolet) limit $\La\to\infty$, then in the far UV $\Gamma$ will closely approximate $\Gamma_0$ and the derivative expansion of $\partial_\La\Gamma$ will closely approximate that of the one-loop correction. On the other hand if we differentiate the flow equation with respect to $\La$ then this results in 
\besp 
\label{ERG2La}
\partial^2_\Lambda \Gamma = +\frac12 \partial_\La |_{\Gamma}\, \text{tr}\left[ \frac{K_\La}{\Delta_{IR}}\left(1+\Delta_{IR}\Gamma^{(2)}\right)^{-1}\right] \\
-\frac12\text{tr}\left[ \frac{K_\La}{\Delta_{IR}}\left(1+\Delta_{IR}\Gamma^{(2)}\right)^{-1}\!\!\Delta_{IR}\,\partial_\La\Gamma^{(2)}\left(1+\Delta_{IR}\Gamma^{(2)}\right)^{-1}\right]  \,,
\eesp
where the first term on the right-hand side differentiates the one-loop-like correction with respect to the cutoff while holding $\Gamma$ constant, whilst the second term adds back the result of differentiating $\Gamma$. This second term takes the form of the two-loop correction \eqref{ERGtwo}. Thus we see that the derivative expansion of $\partial_\La^2\Gamma$ can converge at best like the sum of the one-loop and two-loop corrections, and indeed will approximate this closely in the far UV in the above flow. Iterating in a similar fashion to higher loops, we confirm in this way that non-perturbatively the derivative expansion cannot converge better than what we find by analysing it in the loop expansion.

We identify the dominant Feynman diagram contributions at one loop, two loops and three loops when the derivative expansion is taken to high order. Then, by recognising that the corresponding Schwinger integrals simplify substantially in the limit $n\to\infty$, we obtain analytic expressions for the $O(\partial^{2n})$ contributions at leading order in a large $n$ expansion. Using the same techniques in secs. \ref{sec:othertwo} and \ref{sec:otherthree}, we also confirm that other Feynman diagram contributions are subdominant at large $n$.

At one-loop order the derivative expansion converges. However already at two loops, these results imply that the derivative expansion does not converge. It follows that the derivative expansion cannot converge non-perturbatively either, and furthermore, the very popular approximation technique of expanding the FRG in ever larger sets of local operators, also cannot converge, since this is just an expansion over a subset of all the operators treated in the derivative expansion. 

The leading order contributions give a consistent picture at two and three loops, as follows. At two loops the two-point and four-point $2r$-derivative operators have convergent derivative expansions in any dimension $d$ and for any $r$. The six-point $2r$ operator however only has a convergent derivative expansion if $d<10-2r$. At $d=10-2r$ its derivative expansion oscillates at fixed amplitude as $n\to\infty$ and thus does not converge, while for $d>10-2r$ its derivative expansion diverges, and not in a good way. This means for example we can only get accurate results for the effective potential and thus the equation of state in dimensions $d<10$. On the other hand as we saw in sec. \ref{sec:eightpoint}, if we stay within $d<10-2r$, then the eight-point and all higher point vertices have derivative expansions that diverge in a good way, as oscillating asymptotic series, with convergence up to high order, at least up to $O(\partial^{18})$ for the effective potential in $d=3$ or $4$ dimensions for example, before the derivative expansions then diverge. Then we argue in secs. \ref{sec:threeloop}, \ref{sec:otherthree}, that at high order in the derivative expansion, the three loop contribution of fig. \ref{fig:threeloop} gives the dominant contribution overall to the effective potential. We show that it implies that the derivative expansion is divergent for all monomials $\ph^{2m}$, but for $d\le9$ the leading order derivative expansion is again an oscillatory series of asymptotic type, converging at least up to $O(\partial^{10})$ in $d=3$ dimensions. 

Of course one should keep in mind that these leading order $O(\partial^{2n})$ results are derived in the large $n$ limit. At finite $n$ they are not the full result. But if we formally treat the series on its own then we saw in sec. \ref{sec:asymptotics} that the corresponding exact result always lies between successive partial sums of this series and thus, barring numerical flukes, the most accurate result from summing the leading order derivative expansion is obtained by summing the series to just short of the smallest term. It is interesting then that this oscillatory behaviour, with the exact result being bracketed in this way, appears to be borne out by the highest derivative expansion computation to date, namely $O(\partial^6)$ in ref. \cite{Balog:2019rrg}, except that at $O(\partial^6)$ the assumed best result in the literature actually lies just slightly beyond $O(\partial^6)$ -- outside the range between that and  $O(\partial^4)$. 

However such a direct comparison is not entirely justified also because in ref. \cite{Balog:2019rrg} a parameter $\alpha$ was incorporated in the cutoffs and varied order by order. We treated these more general cutoffs in sec. \ref{sec:gen} where we showed that the large $n$ behaviour is then worse than that with exponential cutoff but again asymptotic of useful type, in fact the asymptotic form is then of the standard $n!\, g^n$ form typically discussed in the literature (the small parameter $g$ here being roughly the inverse distance to the closest singularity in the IR cutoff  propagator). Note that these authors distinguish two types of derivative expansion \cite{Balog:2019rrg,DePolsi:2020pjk}, one that is commonly employed in the literature where different vertices keep their full derivative expansion up to the working order $O(\partial^{2n})$, even when, say, two of them are multiplied together in the flow equation and thus involve expansions for the product that actually extend up to $O(\partial^{4n})$, and a strict version in which one always keeps to $O(\partial^{2n})$ overall, which they follow and in practice they find is better behaved. This distinction matters only for the example in sec. \ref{sec:otherthree}, where indeed we apply the latter ``strict'' version.

The asymptotic series we derive for the leading large $n$ behaviour, using the exponential cutoff \eqref{Rspecial}, have a less severe form, namely that of an expansion of a polylogarithm  outside its radius of convergence, \cf \eqref{polpow}. (At two loops for the two-point and four-point vertex it is the same expansion but inside its radius of convergence.) For a $2m$-point vertex the minimum term lies at what we called the critical value, $n=n_{cr}(m)$. If the large $n$ approximation is good enough at this value, then we would conclude that the full derivative expansion (not just its leading large $n$ part) should be summed to the $(n_{cr}(m)-1)^\text{th}$ term. However we saw in secs. \ref{sec:eightpoint} and \ref{sec:threeloop} that for large $m$ we cannot trust this calculation because $m/n$ corrections, which formally are subleading, in practice are too large around $n\approx n_{cr}(m)$. 

There is also the possibility that, even for low $m$ vertices, other Feynman diagram corrections, which are subleading as $n\to\infty$, could contribute significantly at low $n$ in such a way as to change the picture there, for example to create a minimum term where there was none at leading order (\eg for the six-point vertex in $d\ge 10-2r$ as deduced at two loops). Thus, although we know for sure that the derivative expansion is divergent, a full confirmation that it is an asymptotic expansion with a smallest term at some high order, and under what conditions (\eg on $d$ or $r$), requires going beyond these leading order calculations, possibly to a full non-perturbative treatment.

In this respect one of the strongest pieces of evidence so far is the good convergence already seen in the non-perturbative computations done up to $O(\partial^6)$ \cite{Balog:2019rrg}. 
There, see also \cite{DePolsi:2020pjk,Dupuis:2020fhh}, it has been argued that the derivative expansion is a convergent series, using the observation that the radius of convergence $p^2/m^2=4$ or $9$ of the two-point vertex \eqref{selfe}, as follows by unitarity, where $m$ is the mass in the broken or symmetric phase respectively.\footnote{On the other hand ref. \cite{Golner:1998sr}, see also the review \cite{Bagnuls:2000},  speculated that the derivative expansion could be asymptotic, based on what is seen for a derivative expansion of $QED_{2+1}$ in an external field \cite{Dunne:1996bm,Dunne:1999uy}.} However, the analytic structure of higher-point vertices are also important and more involved, for example already for the three-point vertex  there are so-called anomalous thresholds \cite{Karplus:1958zz} that lie closer to the origin in the complex plane than the singularities demanded by unitarity. We have also seen in sec. \ref{sec:gen} that IR cutoff propagators have more complicated singularities than the pure massive propagator. But the most important mechanism behind the divergence of the derivative expansion, is the factorial dependence on $n$ that is generated when integrating $q^{2n}$ against the cutoff $K_\La(q)$ at order $O(\partial^{2n})$, as in \eqref{momind} (or \eqref{momint} in $d=4$ dimensions) for the exponential cutoff \eqref{Rspecial}, and in \eqref{momgen} for general cutoffs. 

In summary, we have shown that the derivative expansion diverges. We have however also uncovered multiple pieces of evidence that support the conclusion that it is an asymptotic series 
in a regime where accurate results can nevertheless be extracted, despite the fact that there is no small control parameter.

\section*{Acknowledgements}
The author acknowledges support from STFC through Consolidated Grant ST/T000775/1, and thanks Chris J. Howls for enlightening discussions about asymptotic series.


\appendix 

\section{Comparison of derivations}
\label{app:compold}

Here we briefly compare the way the results were derived in sec. \ref{sec:exp}  to how they were derived in ref. \cite{Morris:1999ba}. We also correct for some typos in that reference. There is exact agreement on the expansion for the anomalous dimension, after correcting an error in the overall sign in ref. \cite{Morris:1999ba} eqn. (5.13).\footnote{The equation numbers for that reference refer to the published JHEP article not the preprint version.} There is also exact agreement on the results (\ref{lambdatwofirstterm},\ref{la2ter2deriv}) for the first two terms in \eqref{lambdaflowtwo} contributing to the two-loop flow of $\lambda$. The final term, for which we obtained the expansion \eqref{la2terlast}, does not in fact agree with the expansion found in ref. \cite{Morris:1999ba} because of differences in the way the integrals were evaluated. In particular there it was split into two parts because the derivative expansion was performed on the one-loop six-point sub-diagram before its flow over $\Lambda$ was integrated. One of those parts gave an expression which contains the derivative expansion sum in \eqref{la2terlast} (actually exactly minus this) whilst the other part gave an expansion of the form in \eqref{sa} which, as we have seen, arises from inverting the order of summations. Of course (since the final answer is the same), these two parts together resummed agree with \eqref{lastpartresummed}, although in ref. \cite{Morris:1999ba} we did not recognise that \eqref{la2terlast} could be resummed exactly, as above, so only performed this numerically. In ref. \cite{Morris:1999ba} it was noted that 
the part related to \eqref{sa} had radius of convergence $\Delta=3/2$, whilst it was reported without proof that the $n^\text{th}$ term in the series in \eqref{la2terlast} fell faster than $(2/3)^n/n$ (and thus has at least a radius of convergence of $\Delta=3/2$), as we discussed in sec. \ref{sec:beta}.

In ref. \cite{Morris:1999ba} it was claimed that the two-loop contributions for two-point and four-point vertices when expanded in powers $p^{2r}$ in external momenta, converge just as fast in the $O(\partial^{2n})$ expansion as the zero momentum contributions. Here we found all vertices to converge more slowly by a factor $n^r$, in particular as in \eqref{twlotwint}. In ref. \cite{Morris:1999ba} the claim was based on the observation that the integral \eqref{momint} for $q^{2n-2r}$ would incompletely cancel the $1/n!$ from the one-loop Taylor expansion leaving $\sim 1/n^{2r}$. However after the incomplete cancellation the correct remainder is $\sim 1/n^r$. After this correction the earlier paper is in agreement with the discussion below \eqref{twlotwint}. 

We also found the following typos in ref. \cite{Morris:1999ba}. In its eqn. (5.9), the second line should end in a $+$, not a $\times$. In its (5.10), the first $\ln \frac43$ should appear as $-\ln \frac43$. (5.11) sums to $-12(\lambda^3/[\cdots]$, not $12(\lambda^3/[\cdots]$ as stated. With the sign corrected, (5.13) now equates to $-1/6(\lambda^2[\cdots]$. Finally in (5.14), the first line should have $3\ln\frac43$ (the factor $3$ is missing), and the second line should have $+\frac1{12}\left(\frac12\right)^n$ not $-\frac1{12}\left(\frac12\right)^n$ as stated.

\bibliographystyle{hunsrt}
\bibliography{references}


\end{document}